\documentclass[a4paper,pre,twocolumn,showpacs,aps,floats,superscriptaddress]{revtex4}
\usepackage{epsfig}
\begin{document}

\title{Large-scale topological and dynamical properties of Internet}

\author{Alexei V{\'a}zquez}
\affiliation{International School for Advanced Studies SISSA/ISAS,
  via Beirut 4, 34014 Trieste, Italy}
\author{Romualdo Pastor-Satorras}
\affiliation{Departament de F{\'\i}sica i Enginyeria Nuclear,
  Universitat Polit{\`e}cnica de Catalunya, Campus Nord, M\`{o}dul B4,
  08034 Barcelona, Spain} 
\author{Alessandro Vespignani}
\affiliation{The Abdus Salam International Centre for Theoretical Physics
  (ICTP), P.O. Box 586, 34100 Trieste, Italy}

\date{\today}

\begin{abstract}
  We study the large-scale topological and dynamical properties of
  real Internet maps at the autonomous system level, collected in a 
  three years time interval. We find
  that the connectivity structure of the Internet presents average
  quantities and statistical distributions settled in a well-defined
  stationary state. The large-scale properties are characterized by a
  scale-free topology consistent with previous observations.
  Correlation functions and clustering coefficients exhibit a
  remarkable structure due to the underlying hierarchical organization
  of the Internet. The study of the Internet time evolution shows a
  growth dynamics with aging features typical of recently proposed
  growing network models. We compare the properties of growing network
  models with the present real Internet data analysis.

\end{abstract}


\maketitle

\section{Introduction}
\label{sec:intro}

The Internet is a capital example of growing complex network
\cite{strog01,amaral} interconnecting millions of computers around the
world.  Growing networks exhibit a high degree of wiring entanglement
which takes place during their dynamical evolution. This feature, at
the heart of the new and interesting topological properties recently
observed in growing network systems \cite{barabasi01,dorogorev}, has
triggered the attention of the research community to the study of the
large-scale properties of router-level maps of the Internet
\cite{nlanr,caida,lucent}.  The statistical analysis performed so far
has focused on several quantities exhibiting non-trivial properties:
wiring redundancy and  clustering,
\cite{govindan97,pansiot98,falou99,chou00}, the distribution of
chemical distances \cite{nlanr,falou99}, and the eigenvalue spectra of
the connectivity matrix \cite{falou99}.  Noteworthy, the presence of a
power-law connectivity distribution
\cite{govindan97,falou99,chou00,calda00,yook01} makes the Internet an
example of the recently identified class of scale-free networks
\cite{barab99,barab992}. This evidence implies the absence of any
characteristic connectivity---large connectivity fluctuations---and a
high heterogeneity of the network structure.

As widely pointed out in the literature~\cite{doar93,paxson97,yook01},
a deeper empirical understanding of the topological properties of
Internet is  fundamental in the developing of realistic Internet map
generators, that on their turn are used to test and optimize Internet
protocols.  In fact, the Internet topology has a great influence on
the dynamics that data traffic carries out on top of it.  Hence, a
better understanding of the Internet structure is of primary
importance in the design of new routing \cite{doar93,paxson97} and
searching algorithms \cite{adamic01,puniyani01}, and to protect from
virus spreading \cite{pv01a} and node failures
\cite{barabasi00,newman00,havlin01}.  In this perspective, the direct
measurement and statistical characterization of real Internet maps are
of crucial importance in the identification of the basic mechanisms
that rule the Internet structure and dynamics.

In this work, we shall consider the evolution of real Internet maps
from 1997 to 2000, collected by the National Laboratory for Applied
Network Research (NLANR) \cite{nlanr}, in order to study the
underlying dynamical processes leading to the Internet structure and
topology.  We provide a statistical analysis of several average
properties.  In particular, we consider the average connectivity,
clustering coefficient, chemical distance, and betweenness.  These
quantities will provide a preliminary test of the stationarity of the
network. The scale-free nature of the Internet has been pointed out by
inspecting the connectivity probability distribution, and it implies
that the fluctuations around the average connectivity are not bounded.
In order to provide a full characterization of the scale-free
properties of the Internet, we analyze the connectivity and
betweenness probability distributions for different time snapshot of
the Internet maps. We observe that these distributions exhibit an
algebraic behavior and are characterized by scaling exponents which
are stationary in time. The chemical distance between pairs of nodes,
on the other hand, appears to be sharply peaked around its average
value, providing a striking evidence for the presence of well-defined
small-world properties \cite{watts98}. A more detailed picture of the
Internet can be achieved by studying higher order correlation
functions of the network. In this sense, we show that the Internet
hierarchical structure is reflected in non-trivial scale-free
betweenness and connectivity correlation functions.  Finally, we study
several quantities related to the growth dynamics of the network.  The
analysis points out the presence of two distinct wiring processes: the
first concerns newly added nodes, while the second is related to
already existing nodes increasing their interconnections.  We confirm
that newly added nodes establish new links with the linear
preferential attachment rule often used in modeling growing networks
\cite{barab99}.  In addition, a study of the connectivity evolution of
a single node shows a rich dynamical behavior with aging properties.
The present study could provide some hints for a more realistic
modeling of the Internet evolution, and with this purpose in mind we
provide a discussion of some of the existing growing network models in
the light of our findings. A short account of these results appeared 
in Ref.~\cite{alexei}.

The paper is organized as follows. In Section~\ref{sec:map} we
describe the Internet maps used in our study.  Sec.~\ref{sec:ave} is
devoted to the study of average quantities as a function of time.  In
Sec.~\ref{sec:sta} we provide the analysis of the statistical
distributions characterizing the Internet topology. We obtain evidence
for the scale-free nature of this network as well as for the
stationarity in time of this property.  In Sec.~\ref{sec:hie} we
characterize the hierarchical structure of the Internet by the
statistical analysis of the betweenness and connectivity correlation
functions.  Sec.~\ref{sec:dyn} reports the study of dynamical
properties such as the preferential attachment and the evolution of
the average connectivity of newly added nodes. These properties, which
show aging features, are the basis for the developing of Internet
dynamical models.  Sec.~\ref{sec:model} is devoted to a detailed
discussion of some Internet models as compared
with the presented real data analysis.  Finally, in Sec.~\ref{sec:sum}
we draw our conclusions and perspectives.

\section{Mapping the Internet}
\label{sec:map}

Several Internet mapping projects are currently devoted to obtain
high-quality router-level maps of the Internet.  In most cases, the
map is constructed by using a hop-limited probe (such as the UNIX {\em
  traceroute} tool) from a single location in the network. In this
case the result is a ``directed'', map as seen from a specific
location on the Internet \cite{lucent}.  This approach does not
correspond to a complete map of the Internet because cross-links and
other technical problems (such as multiple Internet provider aliases)
are not considered. Heuristic methods to take into account these
problems have been proposed (see for instance Ref.~\cite{mercator}).
However, it is not clear their reliability and the corresponding
completeness of the maps constructed this way.

A different representation of Internet is obtained by mapping the
autonomous systems (AS) topology.  The Internet can be considered as a
collection of subnetworks that are connected together.  Within each
subnetwork the information is routed using an internal algorithm that
may differ from one subnetwork to another. Thus, each subnet is an
independent unit of the Internet and it is often referred as an AS.
These AS communicate between them using a specific routing algorithm,
the Border Gateway Protocol. Each AS number approximately maps to an
Internet Service Provider (ISP) and their links are inter-ISP
connections. In this case it is possible to collect data from several
probing stations to obtain complete interconnectivity maps (see
Refs.~\cite{nlanr,caida} for a technical description of these
projects).  In particular, the NLANR project is collecting data since
Nov. 1997, and it provides topological as well as dynamical
information on a consistent subset of the Internet. The first Nov.
1997 map contains 3180 AS, and it has grown in time until the Dec.
1999 measurement, consisting of 6374 AS.  In the following we will
consider the graph whose nodes represent the AS and whose links
represent the adjacencies (interconnections) between AS. In particular
we will focus in three different snapshots corresponding to November
8th 1997, 1998, and 1999, that will be referenced as AS97, AS98, and
AS99, respectively.

The NLANR connectivity maps are collected with a resolution of one day
and are changing from day to day. These changes are due to the
addition (birth) and deletion (death) of nodes and links, but also to
the flickering of connections, so that a node may appear to be
isolated (not mapped) from time to time.  A simple test, however,
shows that flickering is appreciable just in nodes with low
connectivity. We compute the ratio $r$ between the number of days in
which a node is observed in the NLANR maps and the total number of
days after the first appearance of the node, averaged over all nodes
in the maps.  The analysis reveals that $r\simeq 1$ and $r>0.65$ for
nodes with connectivity $k\geq 10$, and $k<10$, respectively.  Hence,
nodes with $k<10$ have fluctuations that must be taken into account.
In order to shed light on this point, we inspect the incidence 
of death events with respect to the creation of new nodes.  We consider 
a death event only if a node is not observed in the map during a one 
year time interval. In Table~\ref{tab:0} we show the total number of 
death events in a year, for 1997, 1998, and 1999, in comparison with 
the total number of new nodes created.  
It can be seen that the AS's birth rate appears to be
larger by a factor of two than the death rate. More interestingly, if
we restrict the analysis to nodes with connectivity $k>10$, the death
rate is reduced to a few percent of the birth rate.  This clearly
indicates that only poorly connected nodes have an appreciable
probability to disappear.  This fact is easily understandable in terms
of the market competition among ISP's, where small newcomers are the
ones which more likely go out of business.

\begin{table}[t]
\begin{ruledtabular}
\begin{tabular}{|c|c|c|c|}
Year & 1997 & 1998 & 1999\\
\hline
$N_{\rm new}$ & 309 & 1990 & 3410\\
$N_{\rm dead}$ & 129 & 887 & 1713\\
$N_{\rm dead}(k>10)$ & 0 & 14 & 68\\
\end{tabular}
\caption{Total number of new ($N_{\rm new}$) and deleted ($N_{\rm
    dead}$) nodes 
 in the years 1997, 1998, and 1999.  
 We also report the number of deleted  
 nodes with  connectivity $k>10$.}
\label{tab:0}
\end{ruledtabular}
\end{table}

\section{Average properties and stationarity}
\label{sec:ave}

The growth rate of AS maps reveals that the Internet is a rapidly
evolving network.  Thus, it is extremely important to know whether or
not it has reached a stationary state whose average properties are
time-independent. This will imply that, despite the continuous increase
of nodes and connections in the system, the network's topological
properties are not appreciably changing in time. As a first step, 
we have analyzed the behavior in time of several average 
magnitudes: the average connectivity $\left<k\right>$, 
the clustering coefficient $\left< c\right>$, the average chemical 
distance $\left<d\right>$, and the average betweenness $\left<b\right>$.

The connectivity $k_i$ of a node $i$ is defined as the number of
connections of this node with other nodes in the network, and
$\left<k\right>$ is the average of $k_i$ over all nodes in the
network.  Since each connection contributes to the connectivity of two
nodes, we have that $\left<k\right>=2E/N$, where $E$ is the total
number of connections and $N$ is the number of nodes. Both $E$ and $N$
are increasing with time but their ratio remains almost constant.  The
average connectivity for the 1997, 1998, and 1999 years (averaged over
all the AS maps available for that year) is shown in
Table~\ref{tab:1}. In average each node has three to four connections,
which is a small number compared with that of a fully connected
network of the same size ($\left<k\right>=N-1\sim 10^{3}$). The average
connectivity gives information about the number of connections of any
node but not about the overall structure of these connections. More
information can be obtained using the clustering coefficient
introduced in Ref.~\cite{watts98}. The number of neighbors of a node
$i$ is given by its connectivity $k_i$.  On their turn, these
neighbors can be connected among them forming a triangle with node
$i$. The clustering coefficient $c_i$ is then defined as the ratio
between the number of connections among the $k_i$ neighbors of a given
node $i$ and its maximum possible value, $k_i(k_i-1)/2$. The average
clustering coefficient $\left< c\right>$ is the average of $c_i$ over
all nodes in the network. The clustering coefficient thus provides a
measure of how well locally interconnected are the neighbors of any
node.  The maximum value of $\left< c\right>$ is 1, corresponding to a
fully connected network. For random graphs \cite{bollobas}, which are
constructed by connecting nodes at random with a fixed probability
$p$, the clustering coefficient decreases with the network size $N$ as
$ \left< c \right>_{\rm rand} =\left<k\right>/N$.  On the contrary, it
remains constant for regular lattices.  The average clustering
coefficient obtained for the 1997, 1998, and 1999 years is shown in
Table~\ref{tab:1}. As it can be seen, the clustering coefficient of
the AS maps increases slowly with increasing $N$ and takes values
$\left< c\right>\simeq 0.2$, two orders of magnitudes larger than $ \left<
  c \right>_{\rm rand}\simeq 10^{-3}$, corresponding to a random graph
with the same number of nodes.  Therefore, the AS maps are far from
being a random graph, a feature that can be naively understood using
the following argument: In AS maps the connections among nodes are
equivalent, but they are actually characterized by a real space length
corresponding to the actual length of the physical connection between
AS. The larger is this length, the higher the costs of installation
and maintenance of the line, favoring therefore the connection between
nearby nodes. It is thus likely that nodes within the same
geographical region will have a large number of connection among them,
increasing in this way the local clustering coefficient.

\begin{table}[b]
\begin{ruledtabular}
\begin{tabular}{|c|c|c|c|}
Year  & 1997 & 1998 & 1999\\
\hline
$N$ & 3112 & 3834  & 5287\\
$E$ & 5450 & 6990 & 10100\\
$\left\langle k\right\rangle$ & 3.5(1) & 3.6(1) & 3.8(1)\\
$\left\langle c\right\rangle$ & 0.18(3) & 0.21(3) & 0.24(3)\\
$\left\langle d\right\rangle$ & 3.8(1) & 3.8(1) & 3.7(1)\\
$\left\langle b\right\rangle /N$ &  2.4(1)&  2.3(1) & 2.2(1) 
\end{tabular}
\caption{Average properties of the Internet for three different
  years. $N$: number of nodes; $E$: number of connections;
  $\left\langle k\right\rangle$: average connectivity; $\left\langle
    c\right\rangle$: average clustering coefficient; $\left\langle
    d\right\rangle$ average chemical distance; $\left\langle
    b\right\rangle$ average betweenness. Figures in parenthesis indicate
  the statistical uncertainty from averaging the values of the
  corresponding months in each year.}
\label{tab:1}
\end{ruledtabular}
\end{table}

With this reasoning one might be lead to the conclusion that the
Internet topology is close to a regular two-dimensional lattice. The
analysis of the chemical distances between nodes, however, reveals
that this is not the case.  Two nodes $i$ and $j$ are said to be
connected if one can go from node $i$ to $j$ following the connections
in the network. The path from $i$ to $j$ may be not unique and its
distance is given by the number of nodes visited. The average chemical
distance $\left<d\right>$ is defined as the shortest path distance
between two nodes $i$ and $j$, $d_{ij}$, averaged over every pair of
nodes in the network. For regular lattices, $\left<d\right>_D \sim
N^{1/D}$, where $D$ is the spatial dimension.  Hence, if the Internet
could be mapped into a two-dimensional lattice, we should observe
$\left<d\right>_{\rm D=2}\sim N^{1/2} \approx 60$.  However, as it can be
seen from Table~\ref{tab:1}, for the AS maps $\left<d\right>\simeq 3.6 \ll
\left<d\right>_{\rm D=2}$.  The Internet strikingly exhibits what is
known as the ``small-world'' effect \cite{watts98,watts99}: in average
one can go from one node to any other in the system passing through a
very small number of intermediate nodes.  This necessarily implies
that besides the short local connections which contribute to the large
clustering coefficient, there are some hubs and backbones which
connect different regional networks, strongly decreasing the average
chemical distance.  Another measure of this feature is given by the
number of minimal paths that pass by each node.  To go from one node
in the network to another following the shortest path, a sequence of
nodes is visited.  If we do this for every pair of nodes in the
network, there will be a certain number of key nodes that will be
visited more often than others. Such nodes will be of great importance
for the transmission of information along the network. This fact can
be quantitatively measured by means of the betweenness $b_i$, defined
by the total number of shortest paths between any two nodes in the
network that pass thorough the node $i$.  The average betweenness
$\left<b\right>$ is the average value of $b_i$ over all nodes in the
network. The betweenness has been introduced in the analysis of social
network in Ref.~\cite{newman01b} and more recently it has been studied
in scale-free networks, with the name of load \cite{goh01}.  Moreover,
an algorithm to compute the betweenness has been given in
Ref.~\cite{newman01b}.  For a star network the betweenness takes its
maximum value $N(N-1)/2$ at the central node and its minimum value
$N-1$ at the vertices of the star.  The average betweenness of the
three AS maps analyzed here is shown in Table~\ref{tab:1}. Its value
is between $2N$ and $3N$, which is quite small in comparison with its
maximum possible value $N(N-1)\sim 10^7$.

The present analysis makes clear that the Internet is not dominated by
a very few highly connected nodes similarly to star-shaped
architectures.  As well, simple average measurements rule out the
possibility of a random graph structure or a regular grid
architecture. This evidence hints towards a peculiar topology that
will be fully identified by looking at the detailed probability
distributions of several quantities.  Finally, it is important to
stress that despite the network size is more than doubled in the three
years period considered, the average quantities suffer variations of a
few percent (see Table~\ref{tab:1}).  This points out that the system
seems to have reached a fairly well-defined stationary state, as we
shall confirm in the next Section by analyzing the detailed
statistical properties of the Internet.

\section{Fluctuations and scale-free properties}
\label{sec:sta}

\begin{figure}[t]
\centerline{\epsfig{file=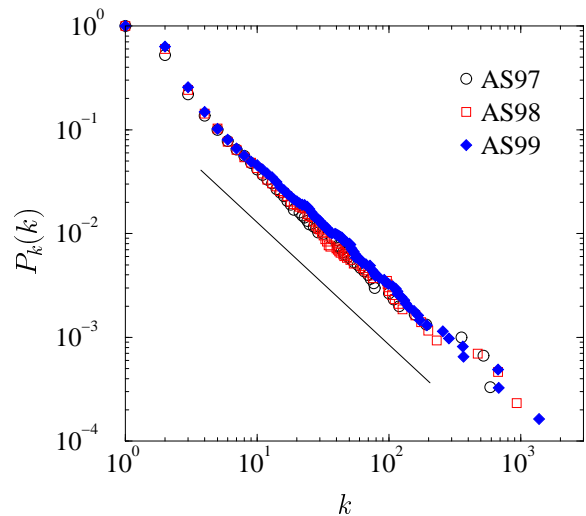,width=3in}}
\caption{Integrated connectivity distribution for the 
  AS97, AS98, and AS99 maps.  The power-law behavior is characterized
  by a slope $-1.2$, which yields a connectivity exponent
  $\gamma=2.2\pm0.1$. }
\label{fig:1}
\end{figure}

In order to get a deeper understanding of the network topology we look
at the probability distributions $p_k(k)$ and $p_b(b)$ that any given
node in the network has a connectivity $k$ and a betweenness $b$,
respectively.  The study of these probability distributions will allow
us to probe the extent of fluctuations and heterogeneity present in
the network.  We shall see that the strong scale-free nature of the
Internet, previously noted in Refs.~\cite{falou99,calda00}, results in
power-law distributions with diverging fluctuations for these
quantities.  The analysis of the maps reveals, in fact, an algebraic
decay for the connectivity distribution,
\begin{equation}
  p_k(k)\sim k^{-\gamma},
\end{equation}
extending over three orders of magnitude.  In Fig.~\ref{fig:1} we
report  the integrated connectivity distribution
\begin{equation}
  P_k(k) = \int_k^\infty p_k(k')dk'
\end{equation}
corresponding to the AS97, AS98, and AS99 maps.  The integrated
distribution, which expresses the probability that a node has
connectivity larger than or equal to $k$, scales as
\begin{equation}
  P_k(k)\sim k^{1-\gamma},
\end{equation}
and it has the advantage of being considerably less noisy that the
original distribution.  In all maps we find a clear power-law behavior
with slope close to $-1.2$ (see Fig.~\ref{fig:1}), yielding a
connectivity exponent $\gamma=2.2\pm 0.1$.  The distribution cut-off
is fixed by the maximum connectivity of the system and is related to
the overall size of the Internet map. We see that for more recent
maps the cut-off is slightly increasing, as expected due to the
Internet growth.  On the other hand, the connectivity exponent
$\gamma$ seems to be independent of time and in good agreement with
previous measurements \cite{falou99}.

\begin{figure*}[t]
\centerline{\epsfig{file=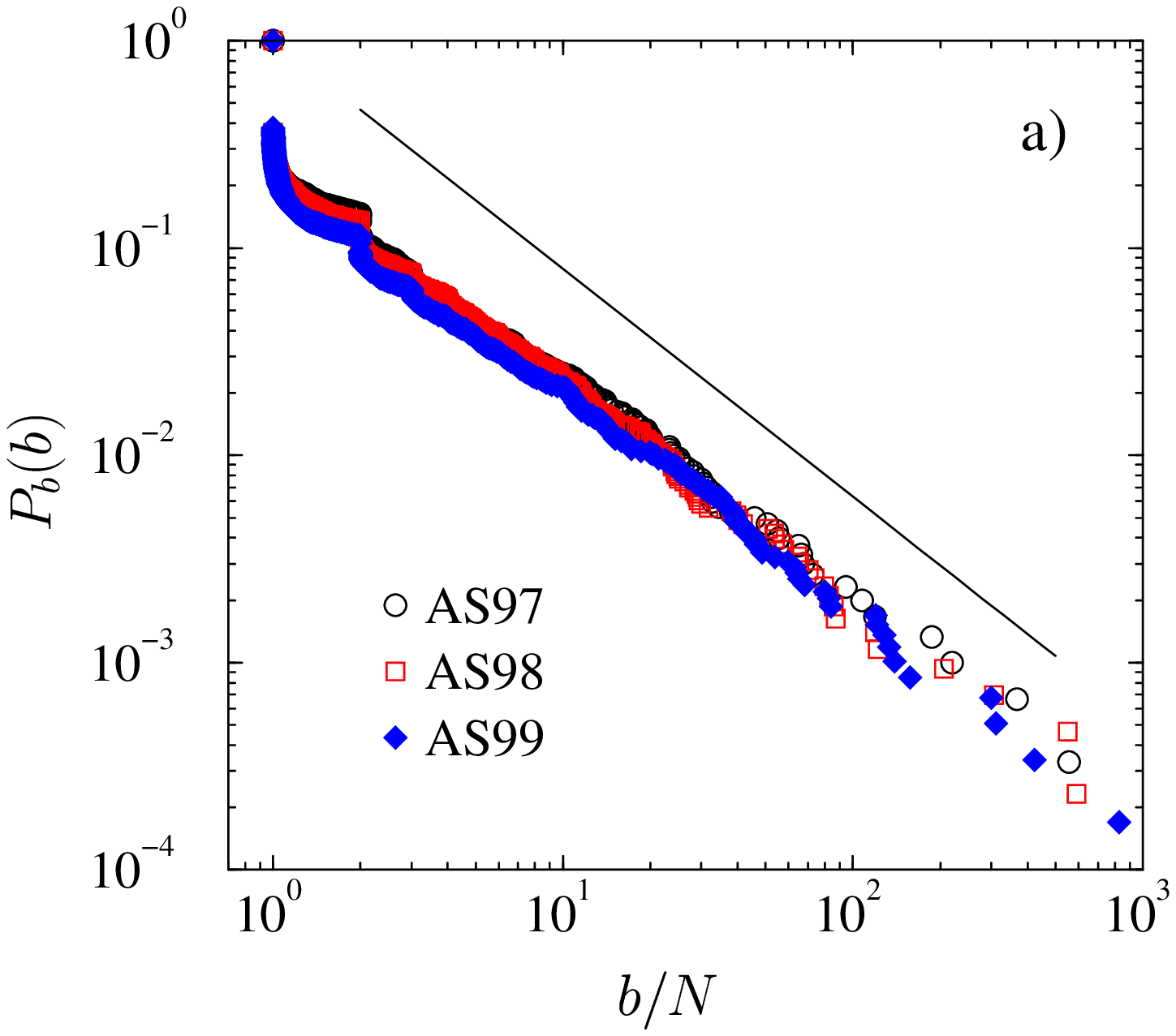,width=3in}
  \epsfig{file=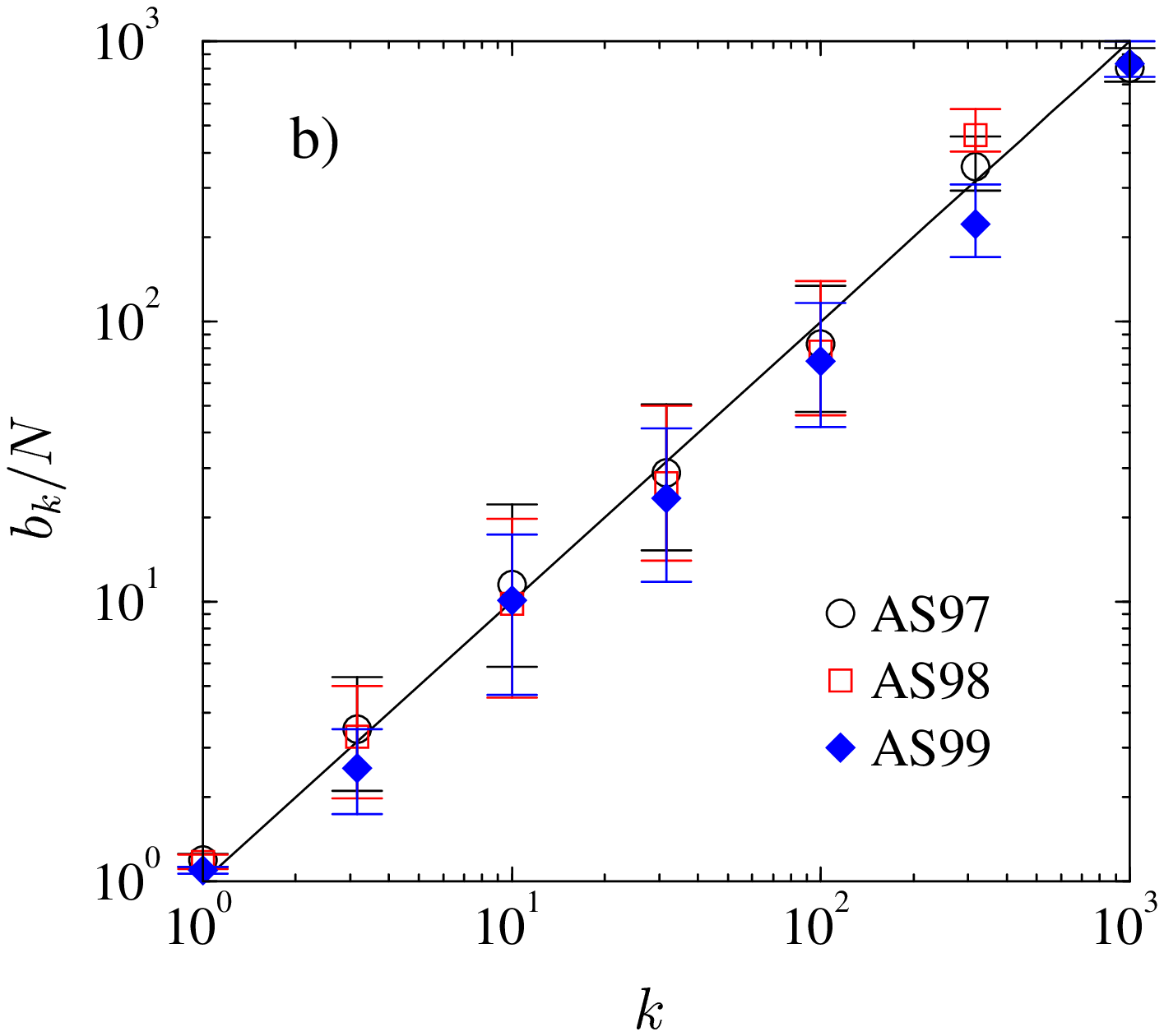,width=3in}} 
\caption{a) Integrated betweenness distribution for the 
  AS97, AS98, and AS99 maps.  The power-law behavior is characterized
  by a slope $-1.1$, which yields a betweenness exponent
  $\delta=2.1\pm0.2$.  b) Betweenness $b_k$ as a function of the
  node's connectivity $k$.  The full line corresponds to the
  predicted behavior $b_k\sim k$. Errors bars take into account statistical 
  fluctuations over different nodes with the same connectivity.}
\label{fig:2}
\end{figure*}

The betweenness distribution $p_b(b)$ (i.e.  the probability that any
given node is passed over by $b$ shortest paths) shows also scale-free
properties, with a a power-law distribution
\begin{equation}
  p_b(b)\sim b^{-\delta}
  \label{eq:between}
\end{equation}
extending over three decades.  As shown in Fig.~\ref{fig:2}(a), the
integrated betweenness distribution measured in the AS maps is 
evidently stable in the three years period analyzed and follows 
a power-law decay
\begin{equation}
P_b(b)= \int_b^\infty p_b(b')db'\sim b^{1-\delta},
\label{eq:intbetween}
\end{equation}
where the  betweenness exponent is $\delta=2.1\pm 0.2$.  
The connectivity and betweenness exponents can be simply 
related if one assumes that the
number of shortest paths $b_k$ passing over a node of connectivity $k$
follows the scaling form 
\begin{equation}
  b_k\sim k^{\beta}.
  \label{eq:b_k}
\end{equation}
By inserting the latter relation in the integrated betweenness 
distribution
Eq.~(\ref{eq:intbetween}) we obtain 
\begin{equation}
  P_k(k)\sim k^{\beta(1-\delta)}.
\end{equation}
Since we have that $P_k(k)\sim k^{1-\gamma}$, we obtain the scaling
relation 
\begin{equation}
  \beta=\frac{\gamma-1}{\delta-1}.
\end{equation}
The measured $\gamma$ and $\delta$ have approximately the same value
for the AS maps data and we expect to recover $\beta\approx 1.0$.
This is corroborated in Fig.~\ref{fig:2}(b), where we report the
direct measurement of the average betweenness of a node as a function
of its connectivity $k$.  It is also worth remarking that it has been
recently argued \cite{goh01} that the betweenness distribution of
scale-free networks with $2<\gamma\leq3$ is an universal quantity not
depending on $\gamma$.  From a numerical study of two scale-free
network models~\cite{goh01}, it was found that the betweenness
distribution follows a universal power-law decay with an exponent
$\delta\approx 2.2$, in fair agreement with our findings.

\begin{figure}[b]
\centerline{\epsfig{file=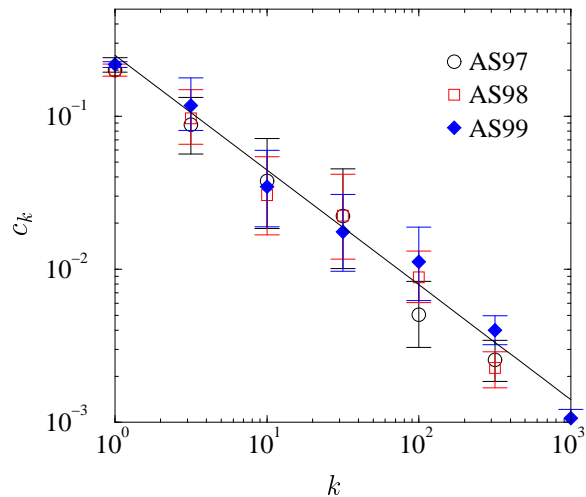,width=3in}}
\caption{Clustering coefficient $c_k$ as a function of the
  connectivity $k$ for the AS97, AS98, and AS99 maps. The best
  fitting  power-law
  behavior is characterized by a slope $-0.75$.Errors bars take into 
  account statistical fluctuations over different nodes with the same connectivity.}
\label{fig:3}
\end{figure}

Another quantity of interest is the probability distribution of the
clustering coefficient of the nodes. In our analysis we
don't find definitive evidence for a power-law behavior of this
distribution. However, still useful information can be gathered from
studying the clustering coefficient $c_k$ as a function of the node
connectivity. In this case the local clustering coefficient of each
node $c_i$ is averaged over all nodes with the same connectivity $k$.
The plots for the AS97, AS98 and AS99 maps are shown in
Fig.~\ref{fig:3}. Also in this case, measurements yield a power-law
behavior $c_k\sim k^{-\omega}$ with $\omega=0.75\pm0.03$, extending
over three orders of magnitudes. This implies that nodes with a small
number of connections have larger local clustering coefficients than
those with a large connectivity.  This behavior is consistent with the
picture previously described in Sec.~\ref{sec:ave} of highly clustered
regional networks sparsely interconnected by national backbones and
international connections.  The regional clusters of AS are probably
formed by a large number of nodes with small connectivity but large
clustering coefficients.  Moreover, they also should contain nodes
with large connectivities that are connected with the other regional
clusters. These large connectivity nodes will be on their turn
connected to nodes in different clusters which are not interconnected
and, therefore, will have a small local clustering coefficient. This
picture also shows the existence of some hierarchy in the network that
will become more evident in the next Section.

\begin{figure*}[t]
\centerline{\epsfig{file=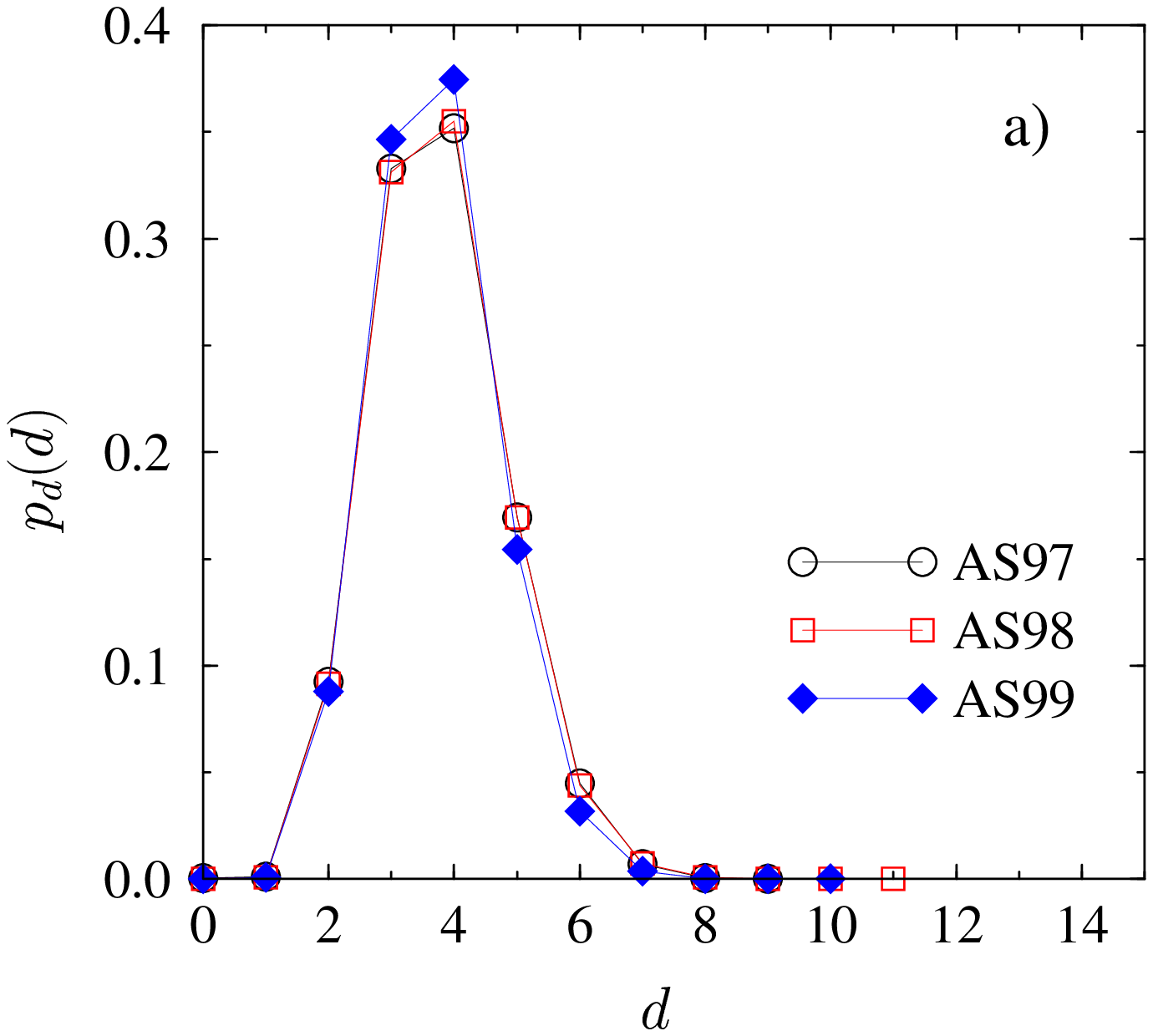,width=3in} \hspace*{0.5cm}
  \epsfig{file=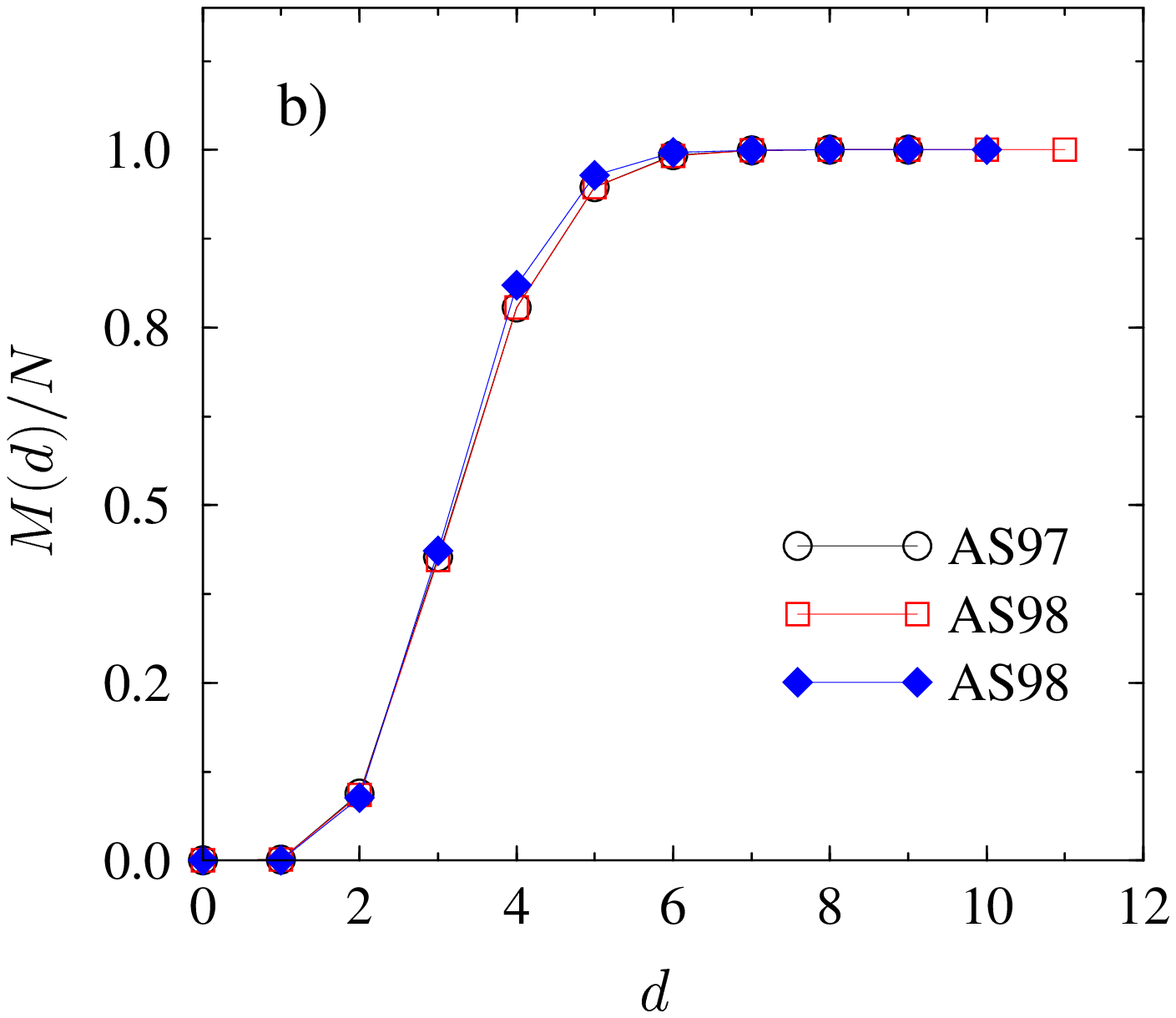,width=3in}}
\caption{a) Distribution of chemical distances $p_d(d)$ for the 
  AS97, AS98, and AS99 maps. b) Hop plots $M(d)$ for the same maps. See 
text for definitions.}
\label{fig:4}
\end{figure*}

A different behavior is followed by the chemical distance $d$ between
two nodes, which does not show singular fluctuations from one pair of
nodes to another. This can be shown by means of the probability
distribution $p_d(d)$ of chemical distances $d$ between pairs of
nodes, reported in Fig.~\ref{fig:4}(a). This distribution is
characterized by a sharp peak around its average value and its shape
remains essentially unchanged from the AS97 to the AS99 maps.
Associated to the chemical distance distribution we have the hop plot
introduced in Ref.~\cite{falou99}.  The hop plot is defined as the
average fraction of nodes $M(d)/N$ within a chemical distance less
than or equal to $d$ from a given node. At $d=0$ we find the starting
node and, therefore, $M(0)=1$. At $d=1$ we found the starting node
plus its neighbors and thus $M(1)=\left<k\right>+1$.  If the network
is made by a single cluster, for $d=d_M$, where $d_M$ is the maximum
chemical distance, $M(d_M)=N$.  For regular $D$-dimensional lattices,
$M(d)\sim d^D$, and in this case $M$ can be interpreted as the mass.
The hop plot is related to the distribution of chemical distances
through the following relation:
\begin{equation}
  \frac{M(d)}{N}=\sum_{d'=0}^d p_d(d').
\end{equation}
The hop plots for the AS97, AS98 and AS99 maps are shown in
Fig.~\ref{fig:4}(b).  In this case the chemical distance barely spans
a decade ($d_M=11$).  Most importantly, $M(d)$ practically reaches its
maximum value $N$ at $d=5$. Hence, the chemical distance does not show
strong fluctuations, as already noticed from the chemical distance
distribution.  In Ref.~\cite{falou99} it was argued that the increase
of $M(d)$ for small $d$ follows a power-law distribution. This
observation is not consistent with the present data, that yield a very
abrupt increase taking place in a very narrow range.

Finally, it is important to stress again that all the measured
distributions are characterized by scaling exponents or behaviors which
are not changing in time. This implies that the statistical properties
characterizing the Internet are time independent, providing a further
test to the network stationarity; i.e. the {\em Internet is
  self-organized in a stationary state characterized by scale-free
  fluctuations}.

\section{Hierarchy and correlations} 
\label{sec:hie}

\begin{figure*}[t]

\centerline{\epsfig{file=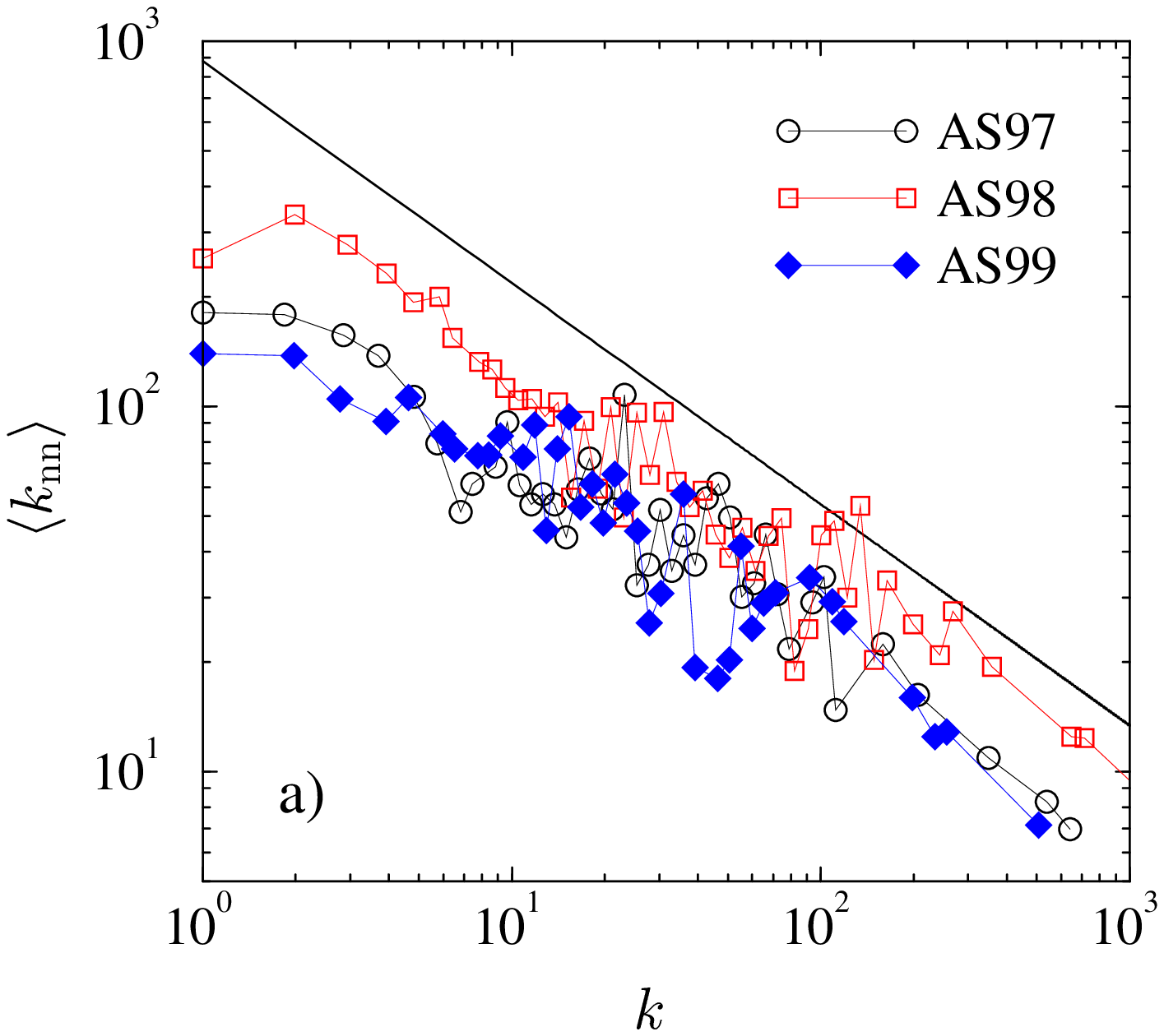, width=3in}
  \epsfig{file=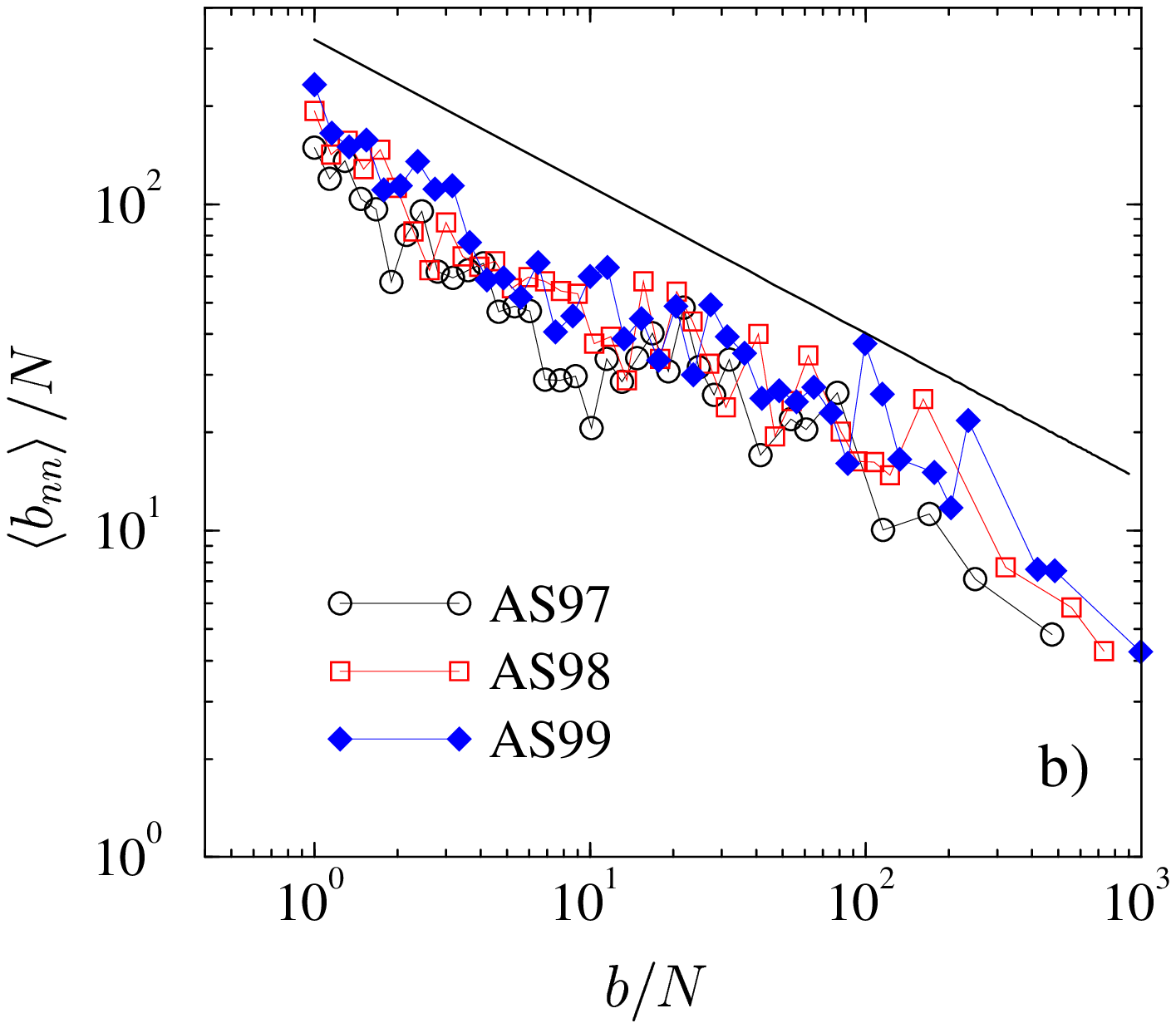, width=3in} }

\caption{a) Average connectivity $\left <k_{nn} \right>$ of the nearest
  neighbors of a node as a function of the connectivity $k$ for the
  same maps. The full line has a slope $-0.5$. b) Average betweenness
  $\left <b_{nn} \right>$ of the nearest neighbors of a node as a
  function of its betweenness $b$ for the AS97, AS98, and AS99 maps.
  The full line has a slope $-0.4$.}
\label{fig:5}
\end{figure*}

Due to installation costs, the Internet has been designed with a
hierarchical structure. The primary known structural difference
between Internet nodes is the distinction between {\em stub} and {\em
transit} domains.  Nodes in stub domains have links that go only
through the domain itself. Stub domains, on the other hand, are
connected via a gateway node to transit domains that, on the contrary,
are fairly well interconnected via many paths. This hierarchy can be
schematically divided into international connections, national
backbones, regional networks, and local area networks.  Nodes
providing access to international connections or national backbones
are of course on top level of this hierarchy, since they make possible
the communication between regional and local area networks.  Moreover,
in this way, a small average chemical distance can be achieved with a
small average connectivity.

Very likely the hierarchical structure will introduce some correlations in the
network topology. We can explore the hierarchical structure of the
Internet by means of the conditional probability $p_c(k'|k)$ that a
link belonging to node with connectivity $k$ points to a node with
connectivity $k'$.  If this conditional probability is independent of
$k$, we are in presence of a topology without any correlation among
the nodes' connectivity. In this case, $p_c(k'| k)=p_c(k')\sim k'
p_k(k')$, in view of the fact that any link points to nodes with a
probability proportional to their connectivity.  On the contrary, the
explicit dependence on $k$ is a signature of non-trivial correlations
among the nodes' connectivity, and the presence of a hierarchical
structure in the network topology.  A direct measurement of the
$p_c(k'| k)$ function is a rather complex task due to large
statistical fluctuations. More clear indications can be extracted by
studying the quantity 
\begin{equation}
  \left<k_{nn}\right>= \sum_{k'}k' p_c(k'| k), 
\end{equation}
i.e.  the nearest neighbors average connectivity of nodes with
connectivity $k$. In Fig.~\ref{fig:5}(a) we show the results obtained
for the AS97, AS98, and AS99 maps, that again exhibit a clear
power-law dependence on the connectivity degree,
\begin{equation}
  \left<k_{nn}\right>\sim k^{-\nu_k},
\end{equation}
with an exponent $\nu_k= 0.5\pm0.1$.  This observation clearly implies
that the connectivity correlation function has a marked dependence
upon $k$, suggesting non-trivial correlation properties for the
Internet.  In practice, this result indicates that highly connected
nodes are more likely pointing to less connected nodes, emphasizing
the presence of a hierarchy in which smaller providers connect to
larger ones and so on, climbing different levels of connectivity.

Similarly, it is expected that nodes with high
betweenness (that is, carrying a heavy load of transit), and
consequently a large connectivity, will be connected to nodes with
smaller betweenness, less load and, therefore, small connectivity. A
simple way to measure this effect is to compute the average
betweenness $\left<b_{nn}\right>$ of the neighbors of the nodes with a
given betweenness $b$. The plot of $\left<b_{nn}\right>$ for the AS97,
AS98, and AS99 maps, represented in Fig.~\ref{fig:5}(b), shows that
the average neighbor betweenness exhibits a clear power-law dependence
on the node betweenness $b$,
\begin{equation}
  \left<b_{nn}\right> \sim b^{-\nu_b},
\end{equation}
with an exponent $\nu_b=0.4\pm0.1$, evidencing that the more loaded
nodes (backbones) are more frequently connected with less loaded nodes
(local networks).

These hierarchical properties of the Internet are likely driven by
several additional factors such as the space locality, economical
resources and the market demand.  An attempt to relate and study some
of these aspects can be found in Ref.~\cite{yook01}, where the
geographical distribution of population and Internet access are
studied. In Sec.~\ref{sec:model} we shall compare a few of the
existing models for the generation of scale-free networks with our
data analysis, in an attempt to identify some relevant features in the
Internet modeling.

\section{Dynamics and growth} 
\label{sec:dyn}

In order to inspect the Internet dynamics, we focus our attention on
the addition of new nodes and links into the maps. In the three-years
range considered, we keep track of the number of links $\ell_{\rm
  new}$ appearing between a newly introduced node and an already
existing node.  We also monitor the rate of appearance of links
$\ell_{\rm old}$ between already existing nodes.  In Table \ref{tab:2}
we can observe that the creation of new links is governed by these two
processes at the same time. Specifically, the largest contribution to
the growth is given by the appearance of links between already
existing nodes.  This clearly points out that the Internet growth is
strongly driven by the need of redundancy in the wiring and an
increased need of available bandwidth for data transmission.

\begin{table}[b]
\begin{ruledtabular}
\begin{tabular}{|c|c|c|c|}
Year & 1997 & 1998 & 1999\\
\hline
$\ell_{\rm new}$ & 183(9) & 170(8) & 231(11)\\
$\ell_{\rm old}$ & 546(35) & 350(9) & 450(29)\\
$\ell_{\rm new}/\ell_{\rm old}$ & 0.34(2) & 0.48(2) & 0.53(3)
\end{tabular}

\caption{Monthly rate of new links connecting existing 
nodes to new ($\ell_{\rm new}$) and old ($\ell_{\rm old}$)
nodes.}

\label{tab:2}
\end{ruledtabular}
\end{table}

\begin{figure}[t]

\centerline{\epsfig{file=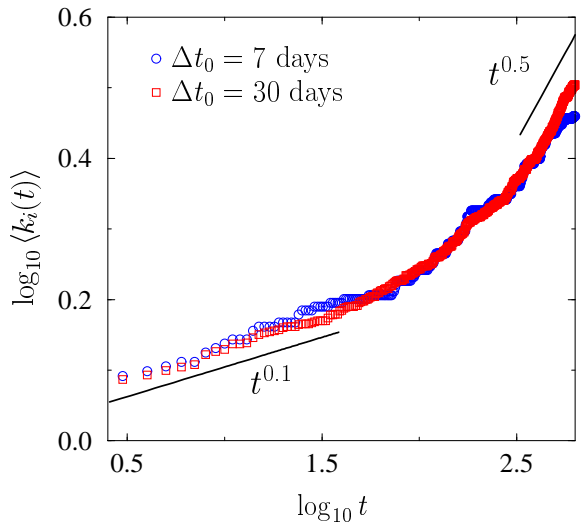, width=3in}}

\caption{Average connectivity of nodes borne within a small time
    window $\Delta t_0$, after a time $t$ elapsed since their appearance. 
    Time $t$ is measured in days.  As a comparison we report the lines
    corresponding to $t^{0.1}$ and $t^{0.5}$.}

\label{fig:6}
\end{figure}

A customarily measured quantity in the case of growing networks is the
average connectivity $\left<k_i(t)\right>$ of new nodes as a function
of their age $t$. In Refs.~\cite{barab992,krap00,krap00b} it is shown that
$\left<k_i(t)\right>$ is a scaling function of both $t$ and the
absolute time of birth of the node $t_0$. We thus consider the total
number of nodes born within an small observation window $\Delta t_0$, such
that $t_0\simeq {\rm const.}$ with respect to the absolute time scale that
is the Internet lifetime. For these nodes, we measure the average
connectivity as a function of the time $t$ elapsed since their birth.
The data for two different time windows are reported in
Fig.~\ref{fig:6}, where it is possible to distinguish two different
dynamical regimes: At early times, the connectivity is nearly constant
with a very slow increase ($\left<k_i(t)\right>\sim t^{0.1}$). Later on,
the behavior approaches a power-law growth $\left<k_i(t)\right>\sim
t^{0.5}$.  While exponent estimates are affected by noise and limited
time window effects, the crossover between two distinct dynamical
regimes is compatible with the general aging form obtained in the context of
growing networks  in Ref.~\cite{krap00,krap00b}.  

\begin{figure}[t]

\centerline{\epsfig{file=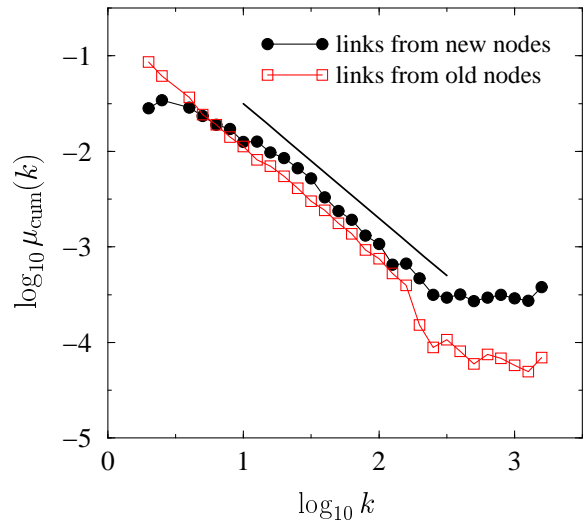, width=3in}}
\caption{Integrated frequency of links emanating 
  from new and existing nodes that attach to nodes with connectivity
  $k$. The full line corresponds to a slope $-0.2$, which yields an
  exponent $\alpha \simeq 1.0$.  The flat tails are originated from
  the poor statistics at very high $k$ values.}

\label{fig:7}

\end{figure}

A very important issue in the modeling of growing networks concerns
the understanding of the growth mechanisms at the origin of the
developing of new links. As we shall see more in detail in the next
Section, the basic ingredients in the modeling of scale-free growing
networks is the preferential attachment hypothesis \cite{barab99}.  In
general, all growing network algorithms define models in which the
rate $\Pi(k)$ with which a node with $k$ connections receives new links
is proportional to $k^\alpha$ (see Ref.~\cite{barab99} and
Sec.~\ref{sec:model}). The inspection of the exact value of $\alpha$ in
real networks is an important issue since the connectivity properties
strongly depend on this exponent~\cite{krap00,krap00b,barabba01b}.
Here we use a simple recipe that allows to extract the value of $\alpha$
by studying the appearance of new links.  We focus on links emanating
from newly appeared nodes in different time windows ranging from one
to three years. We consider the frequency $\mu(k)$ of links that
connect to nodes with connectivity $k$.  By using the preferential
attachment hypothesis, this effective probability is $\mu(k)\sim k^\alpha
p_k(k)$.  Since we know that $p_k(k)\sim k^{-\gamma}$, we expect to find a
power-law behavior $\mu(k)\sim k^{\alpha-\gamma}$ for the frequency. In Fig.
\ref{fig:7} we report the obtained results for the integrated
frequency $\mu_{\rm cum}(k) = \int_k^\infty \mu(k') dk'$, which shows a
behavior compatible with an algebraic dependence $\mu(k)\sim k^{-1.2}$.
By using the independently obtained value $\gamma=2.2$ we find a
preferential attachment exponent $\alpha\simeq 1.0$, in good agreement with
the result obtained with a different analysis in
Ref.~\cite{barabba01b}.  We performed a similar analysis also for
links emanated by existing nodes, recovering the same form of
preferential attachment (see Fig. \ref{fig:7}).  The present analysis
confirms the validity of the preferential attachment hypothesis, but
leaves open the question of the interplay with several other factors,
such as the nodes' hierarchy, space locality, and resource
constraints.

\section{Modeling the Internet}
\label{sec:model}

In the previous Section we have presented a thorough analysis of the
AS maps topology. Apart from providing useful empirical data to
understand the behavior of the Internet, our analysis is of great
relevance in order to test the validity of models of the Internet
topology. The Internet topology has a great influence on the
information traffic carried on top of it, including routing algorithms
\cite{doar93,paxson97}, searching algorithms
\cite{adamic01,puniyani01}, virus spreading \cite{pv01a}, and
resilience to node failure \cite{barabasi00,newman00,havlin01}. Thus,
designing network models which accurately reproduce the Internet
topology is of capital importance to carry out simulations on top of
these networks.

Early works considered the Erd\"{o}s-R\'{e}nyi~\cite{erdos60} model
or hierarchical networks as models of Internet~\cite{zeg}.  However,
they yield connectivity distributions with a fast (exponential) decay
for large connectivities, in disagreement with the power-law decay
observed in real data. Only recently the Internet modeling benefited 
of the  major advance provided in the field of growing  
networks by the introduction of the
Barab\'{a}si-Albert (BA) model \cite{barab99,barab992,mendes99}, which
is related to 1955 Simon's model
\cite{simon55,bornholdt01,dmscomment}.  The main ingredients of this
model are the growing nature of the network and a preferential
attachment rule, in which the probability of establishing new links
toward a given node grows linearly with its connectivity.  
The BA model is constructed using the following algorithm \cite{barab99}: 
We start from a small number $m_0$ of disconnected nodes; every time 
step a new node is added, with $m$ links that are connected to an 
old node $i$ with probability
\begin{equation}
  \Pi_{\rm BA}(k_i) = \frac{k_i}{\sum_j k_j},
  \label{eq:PiBA}
\end{equation}
where $k_i$ is the connectivity of the $i$-th node. After iterating
this procedure $N$ times, we obtain a network with a connectivity
distribution $p_k(k) \sim k^{-3}$ and average connectivity $\left< k
\right> = 2m$. In this model, heavily connected nodes will increase
their connectivity at a larger rate than less connected nodes, a
phenomenon that is known as the ``rich-get-richer'' effect
\cite{barab99}.  It is worth remarking, however, that more general
studies \cite{krap00,krap00b,dorogorev} have revealed that nonlinear
attachment rates of the form $\Pi(k) \sim k^\alpha$ with $\alpha \neq 1$ have as an
outcome connectivity distributions that depart form the power-law
behavior.  The BA model has been successively modified with the
introduction of several ingredients in order to account for
connectivity distribution with $2<\gamma<3$
\cite{albert00,krap00,krap00b}, local geographical
factors~\cite{medina}, wiring among existing nodes~\cite{mendes00},
and age effects~\cite{mendes01}.

In the previous Section we have analyzed different measures that
characterize the structure of AS maps. Since several models are able 
to reproduce the right power law behavior for the connectivity distribution, 
the analysis obtained in the previous sections can provide the effective 
tools to scrutinize the different models at a deeper level.  
In particular, we perform  a data comparison  for three different
models that generate networks with power-law connectivity
distributions. 
First we have considered a random graph constructed
with a power-law connectivity distribution, using the Molloy and Reed
(MR) algorithm \cite{molloy95,molloy98}. Secondly, we have studied two
variations of the BA model, that yield connectivity exponents
compatible with the one measured in Internet: the generalized
Barab\'{a}si-Albert (GBA) model \cite{albert00}, which includes the
possibility of connection rewiring, and the fitness model
\cite{bianconi01}, that implements a weighting of the nodes in the
preferential attachment probability.

\begin{figure}[t]
\centerline{\epsfig{file=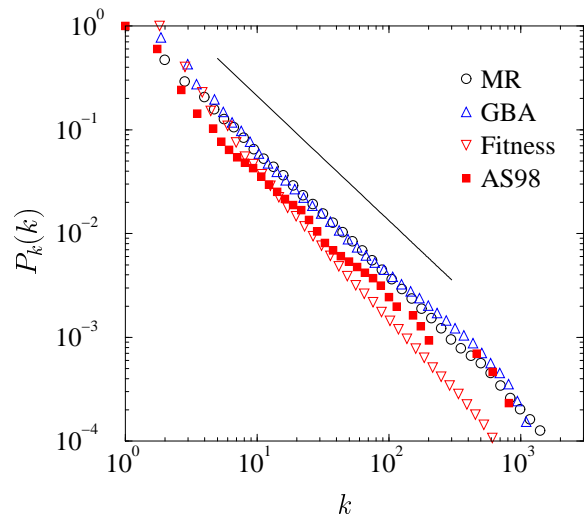,width=3in}}
\caption{Integrated connectivity distribution for the  MR, GBA, and
  fitness models, compared with the result from the AS98 map. The full
  line has slope $-1.2$.}
\label{fig:m0}
\end{figure}

The models are defined as follows:

{\bf MR model}: In the construction of this model
\cite{molloy95,molloy98,newman01c,dorogorev} we start assigning to
each node $i$ in a set of $N$ nodes a random connectivity $k_i$ drawn
from the probability distribution $p_k(k)\sim k^{-\gamma}$, with
$m\leq k_i<N$, and imposing the constraint that the sum $\sum_i k_i$
must be even. The graph is completed by randomly connecting the nodes
with $\sum_i k_i/2$ links, respecting the assigned connectivities.
The results presented here are obtained using $m=1$ and a connectivity
exponent $\gamma=2.2$, equal to that found in the AS maps.
Clearly this construction algorithm does not take into account any 
correlations or dynamical feature of the Internet and it can be considered
as a first order approximations that focuses only in the connectivity
properties.

{\bf GBA model}: It is defined by starting with $m_0$ nodes connected
in a ring~\cite{albert00}: At each time step one of the following
operations is performed:
\begin{itemize}
\item [(i)] With probability $q$ we rewire $m$ links. For each of
  them, we randomly select a node $i$ and a link $l_{ij}$ connected to
  it. This link is  removed and replaced by a new link
  $l_{i^\prime j}$ connecting the node $j$ to a new node $i^\prime$
  selected with probability $\Pi(k_{i^\prime})$ where 
  \begin{equation}
    \Pi_{\rm GBA}(k_i)=\frac{k_i+1}{\sum_j(k_j+1)}.
    \label{eq:m1}
\end{equation}
\item [(ii)] With probability $p$ we add $m$ new links. For each of
  them, one end of the link is selected at random, while the other is
  selected with probability as in Eq.~(\ref{eq:m1}).
\item [(iii)] With probability $1-q-p$ we add a new node with $m$
  links that are connected to nodes already present with probability
  as in Eq.~(\ref{eq:m1}). 
\end{itemize}
The preferential attachment probability Eq.~(\ref{eq:m1}) leads to a
power-law distributed connectivity, whose exponent depends on the
parameters $q$ and $p$. In the particular case $p=0$, the connectivity
exponent is given by \cite{albert00}
\begin{equation}
  \gamma = 1 + \frac{(1-q)(2m+1)}{m}.
  \label{eq:gammaBA}
\end{equation}
Hence, changing the value of $m$ and $q$ we can obtain the desired
connectivity exponent $\gamma$. In the present simulations we use the
values $m=2$ and $q=13/25$, that yield the exponent $\gamma=2.2$.  The GBA
model embeds both the rich-get-richer paradigm and the growing nature
of the Internet; however, it does not take into account any possible
difference or hierarchies in newly appearing nodes.

{\bf Fitness model}: This network model introduces an external
competence among nodes to gain links, that is controlled by a random
(fixed) fitness parameter $\eta_i$ that is assigned to each node $i$
from a probability distribution $\rho(\eta)$.  In this case, we also
start with $m_0$ nodes connected in a ring and at each time step we
add a new node $i'$ with $m$ links that are connected to nodes already
present on the network with probability
\begin{equation}
  \Pi_{\rm fitness}(k_i)=\frac{\eta_ik_i}{\sum_j\eta_jk_j}.
\label{eq:m2}
\end{equation}
The newly added node is assigned a fitness $\eta_{i'}$. The results presented
here are obtained using $m=2$ and a probability $\rho(\eta)$ uniformly
distributed in the interval $[0,1]$, which yields a connectivity
distribution $p_k(k)\sim k^{-\gamma}/\ln k$ with $\gamma \approx 2.26$
\cite{bianconi01}. The fitness model adds to the growing dynamics with
preferential attachment a stochastic parameter, the fitness, that
embeds all the properties, other than the connectivity, that may
influence the probability of gaining new links.

\begin{table}[b]
\begin{ruledtabular}
\begin{tabular}{|c|c|c|c|c|}
 & MR  & GBA  & Fitness  & 1998\\
\hline
$\left\langle k\right\rangle$    & 4.8(1)  & 5.4(1)   & 4.00(1) & 3.6(1)  \\
$\left\langle c\right\rangle$    & 0.16(1) & 0.12(1)  & 0.02(1) & 0.21(3)  \\
$\left\langle d\right\rangle$    & 3.1(1)  & 1.8(1)   & 4.0(1)  & 3.8(1)  \\
$\left\langle b\right\rangle /N$ & 2.2(1)  & 1.9(1)   & 2.1(1)  & 2.3(1)
\end{tabular}
\caption{Average properties of the MR, GBA, and fitness models,
  compared with the values from the Internet in 1998.  $\left\langle
    k\right\rangle$: average connectivity; $\left\langle
    c\right\rangle$: average clustering coefficient; $\left\langle
    d\right\rangle$ average chemical distance; $\left\langle
    b\right\rangle$ average betweenness. Figures in parenthesis
  indicate the statistical uncertainty from the average of $1000$
  realizations of the models.}
\label{tab:3}
\end{ruledtabular}
\end{table}

We have performed simulations of these three models using the
parameters mentioned above and using sizes of $N\simeq 4000$ nodes, in
analogy with the size of the AS map analyzed.  In each case we perform
averages over $1000$ different realizations of the networks. It is
worth remarking that while the fitness model generates a connected
network, both the GBA and the MR model yield disconnected networks.
This is due to the rewiring process in the GBA model, while the
disconnect nature of the graph in the MR model is an inherent
consequence of the connectivity exponent being larger that $2$
\cite{newman01c}. In these two cases we therefore work with graphs
whose giant component (that is, the largest cluster of connected nodes
in the network \cite{bollobas}) has a size of the order $N$. It is
important to remind the reader that we are working with networks of a
relatively small size, chosen so as to fit the size of the Internet
maps analyzed in the previous Sections.  In this perspective, all the
numerical analysis that we shall perform in the following serve only
to check the validity of the models as representations of the Internet
as we know it, and do not refer to the intrinsic properties of the
models in the thermodynamic limit $N\to\infty$.

As a first check of the connectivity properties of the models, in
Fig~\ref{fig:m0} we have plotted the integrated
connectivity distributions. For the MR model we recover the expected
exponent $\gamma_{\rm MR} \simeq 2.20$, since it was imposed in the
very definition of the model.  For the GBA model we obtain numerically
$\gamma_{\rm GBA} \simeq 2.19$ for the giant component, in excellent
agreement with the value predicted by Eq.~(\ref{eq:gammaBA}) for the
asymptotic network.  For the fitness model, on the other hand, 
a numerical regression of the integrated connectivity distribution 
yields an effective exponent $\gamma_{\rm fitness} \simeq 2.4$.  
This value is larger than  the theoretical prediction $2.26$ obtained 
for the model~\cite{bianconi01}.
The discrepancy is mainly due to the logarithmic corrections 
present in the connectivity distribution of this model.
These corrections are more evident in the relatively
small-sized networks used in this work and become progressively
smaller for larger network sizes.

\begin{figure*}[t]
\centerline{\epsfig{file=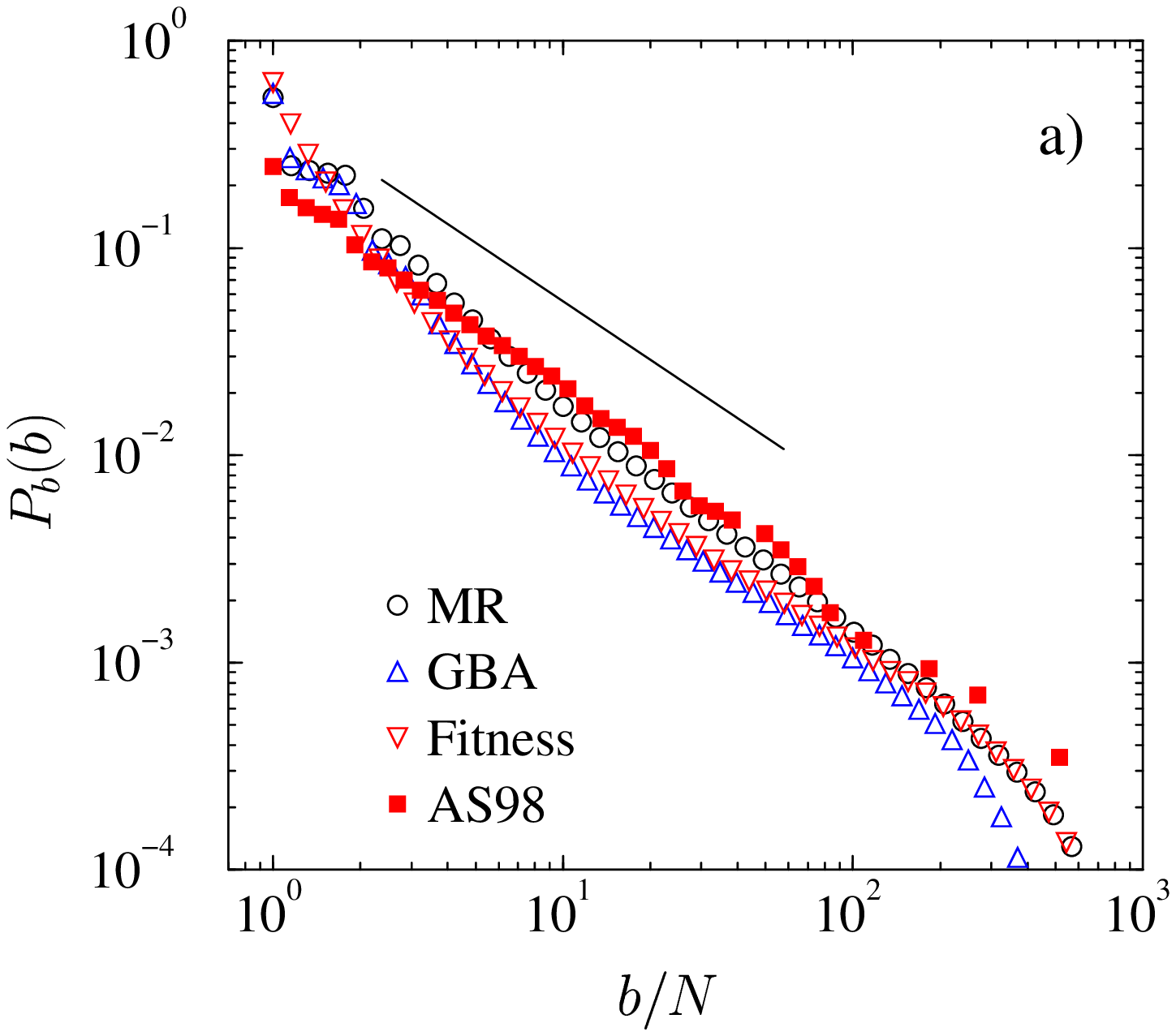,width=3in}
  \epsfig{file=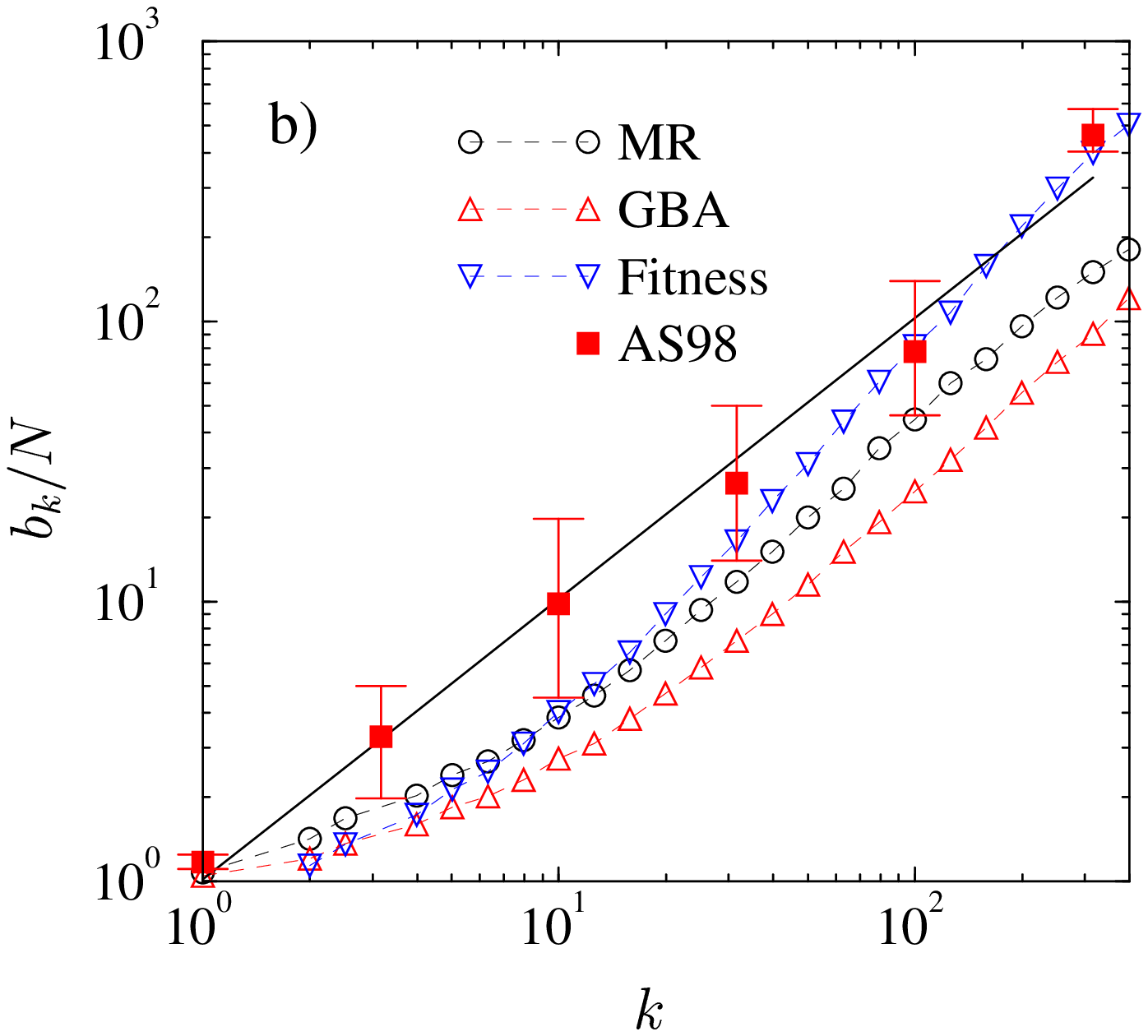,width=3in}}
\caption{a) Integrated betweenness distribution for the  MR, GBA, and
  fitness models, compared with the result from the AS98 map. The full
  line on to has slope $-1.1$, corresponding to the Internet map.  b)
  Betweenness $b_k$ as a function of the node's connectivity $k$
  corresponding to the previous results. The full line has slope $1.0$.}
\label{fig:m1}
\end{figure*}

In Table~\ref{tab:3} we report the average values of the connectivity,
clustering coefficient, chemical distance, and betweenness for the
three models, compared with the respective values computed for
Internet during 1998. From the examination of this Table, one could
surprisingly conclude that the MR model, which neglects by 
constructions any correlation among nodes, yields the average values 
in better agreement with the Internet data. As we can observe, 
the fitness model provides a too small value  for the average 
clustering coefficient, while the GBA model clearly fails for the 
average chemical distance and the betweenness. 
A more crucial test about the models is however provided  
by the analysis of the full distribution of the various quantities,
that should reproduce the scale-free features of the real Internet. 

The betweenness distribution $p_b(b)$ of the three
models give qualitatively similar results. The integrated betweenness
distribution $P_b(b)$ obtained is plotted in
Fig.~\ref{fig:m1}(a). Both the MR and the fitness models follow a
power-law decay $P_b(b)\sim b^{-\delta}$ with an exponent $\delta\simeq
2$, in agreement with the value obtained from the AS maps.  The GBA
model shows an appreciable bending which, nevertheless, is
compatible with the experimental Internet behavior.
These results are in agreement with the numerical prediction in
Ref.~\cite{goh01} and support the conjecture that 
the exponent $\delta\simeq 2.2$ is a universal quantity in 
all scale-free networks with $2<\gamma<3$. 
In order to further inspect the betweenness properties, 
we plot in Fig.~\ref{fig:m1}(b) the average betweenness $b_k$ as 
a function of the connectivity. In this
case, the MR and GBA models yield an exponent $\beta \simeq 1$,
compatible with the AS maps, while the fitness model exhibits a somewhat
larger exponent, close to $1.4$. Also in this case, we have that the 
finite size logarithmic  corrections present in the fitness model could 
play a determinant role in this discrepancy.

\begin{figure}[b]
\centerline{\epsfig{file=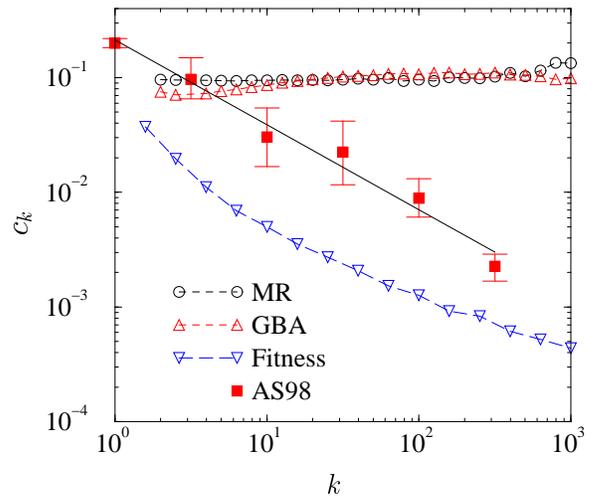,width=3in}}
\caption{Clustering coefficient $c_k$ as a function of the
  connectivity $k$ for the MR, GBA, and fitness models, compared with
  the result from the AS98 map. The full line has slope $-0.75$.}

\label{fig:m2}
\end{figure}

\begin{figure}[b]
\centerline{\epsfig{file=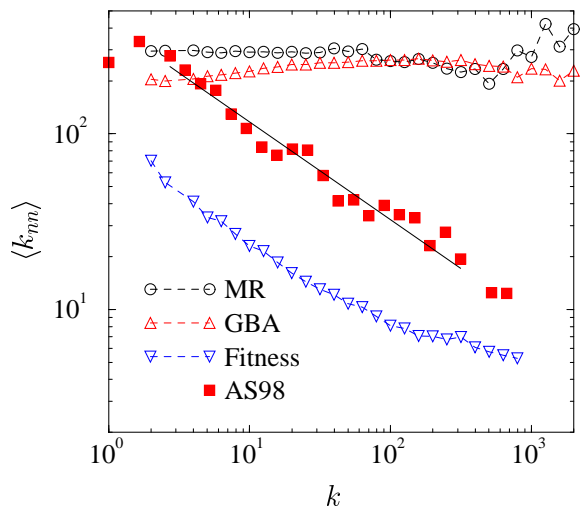,width=3in}}
\caption{Average connectivity  of the nearest
  neighbors of a node as a function of the connectivity $k$ for the
  MR, GBA, and fitness models, compared with the result from the AS98
  map. The AS98 data have been binned for the sake of clarity.
  The full line has a slope $-0.5$.}
\label{fig:m3}
\end{figure}

While properties related to the betweenness do not appear to pinpoint
a major difference among the models, the most striking test is
provided by analyzing the correlation properties of the models.  In
Figs.~\ref{fig:m2} and~\ref{fig:m3}, we report the average clustering
coefficient as a function of the connectivity, $c_k$, and the average
connectivity of the neighbors, $\left< k_{nn} \right>$, respectively.
The data from Internet maps show a nontrivial $k$ structure that, as
discussed in previous Sections, is due to scale-free correlation
properties among nodes. These properties depend on their turn upon the
underlying hierarchy of the Internet structure. The only model that
renders results in qualitative agreement with the Internet maps is the
fitness model.  On the contrary, the MR and GBA models completely
fail, producing quantities which are almost independent on $k$. The
reason of this striking difference can be traced back to the lack of
correlations among nodes, which in the MR model is imposed by
construction (the model is a random network with fixed connectivity
distribution), and in the GBA model it is due to the destruction of
correlations by the random rewiring mechanism implemented. The general
analytic study of connectivity correlations in growing networks models
can be found in Ref.~\cite{krap00b}, and it is worth noticing that a
$k$-structure in correlation functions, as probed by the quantity
$\left<k_{nn}\right>$, does not arise in all growing network models.
In this perspective we can use correlation properties as one of the
discriminating feature among various models that show the same
scale-free connectivity exponent.

The fitness model is able to reproduce the non trivial correlation
properties because of the fitness parameter of each node that mimics
the different hierarchical, economical, and geographical constraints
of the Internet growth. Since the model is embedding many features in
one single parameter, we have to consider it just as a very first step
towards a more realistic modeling of the Internet.  In this
perspective, models in which the attachment rate depends on both the
connectivity and the real space distance between two nodes has been
studied in \cite{medina,yook01}.  These models seem to give a better
description of the Internet topology.  In particular, the model of
Ref.~\cite{yook01}, includes geographical constraints, obtaining that,
on average, the probability to connect to a given node scales linearly
with its connectivity and it is inversely proportional to the distance
to that node.  A comparison with real data is, in this case, more
difficult because Internet maps generally lack geographical and
economical information.

\section{Summary and conclusions}
\label{sec:sum}

In summary, we have shown that the Internet maps exhibit a stationary
scale-free topology, characterized by non-trivial connectivity
correlations. An investigation of the Internet dynamics confirms the
presence of a preferential attachment behaving linearly with the
nodes' connectivity and identifies two different dynamical regimes
during the nodes' evolution.  We have compared several models of
scale-free networks to the experimental data obtained from the AS
maps. While all the models seem to capture the scale-free connectivity
distribution, correlation and clustering properties are captured only
in models that take into account several other ingredients, such as
the nodes' hierarchy, resource constraints and geographical location.
Other ingredients that should be included in the Internet modeling
concern the possibility of including the wiring among existing nodes
and age effects that our analysis show to be an appreciable feature of
the Internet evolution.  The results presented in this work show that
the understanding and modeling of Internet is an interesting and
stimulating problem that need the cooperative efforts of data analysis
and theoretical modeling.

\begin{acknowledgments}
  This work has been partially supported by the European Network
  Contract No. ERBFMRXCT980183.  R.P.-S. acknowledges financial
  support from the Ministerio de Ciencia y Tecnolog\'{\i}a (Spain) and
  from the Abdus Salam International Centre for Theoretical Physics
  (ICTP), where part of this work was done.
\end{acknowledgments}


\begin{thebibliography}{47}
\expandafter\ifx\csname natexlab\endcsname\relax\def\natexlab#1{#1}\fi
\expandafter\ifx\csname bibnamefont\endcsname\relax
  \def\bibnamefont#1{#1}\fi
\expandafter\ifx\csname bibfnamefont\endcsname\relax
  \def\bibfnamefont#1{#1}\fi
\expandafter\ifx\csname citenamefont\endcsname\relax
  \def\citenamefont#1{#1}\fi
\expandafter\ifx\csname url\endcsname\relax
  \def\url#1{\texttt{#1}}\fi
\expandafter\ifx\csname urlprefix\endcsname\relax\def\urlprefix{URL }\fi
\providecommand{\bibinfo}[2]{#2}
\providecommand{\eprint}[2][]{\url{#2}}

\bibitem[{\citenamefont{Strogatz}(2001)}]{strog01}
\bibinfo{author}{\bibfnamefont{S.~H.} \bibnamefont{Strogatz}},
  \bibinfo{journal}{Nature} \textbf{\bibinfo{volume}{410}},
  \bibinfo{pages}{268} (\bibinfo{year}{2001}).

\bibitem[{\citenamefont{Amaral et~al.}(2000)\citenamefont{Amaral, Scala,
  Barth\'{e}l\'{e}my, and Stanley}}]{amaral}
\bibinfo{author}{\bibfnamefont{L.~A.~N.} \bibnamefont{Amaral}},
  \bibinfo{author}{\bibfnamefont{A.}~\bibnamefont{Scala}},
  \bibinfo{author}{\bibfnamefont{M.}~\bibnamefont{Barth\'{e}l\'{e}my}},
  \bibnamefont{and} \bibinfo{author}{\bibfnamefont{H.~E.}
  \bibnamefont{Stanley}}, \bibinfo{journal}{Proc. Natl. Acad. Sci. USA}
  \textbf{\bibinfo{volume}{97}}, \bibinfo{pages}{11149} (\bibinfo{year}{2000}).

\bibitem[{\citenamefont{Albert and Barab\'{a}si}(2001)}]{barabasi01}
\bibinfo{author}{\bibfnamefont{R.}~\bibnamefont{Albert}} \bibnamefont{and}
  \bibinfo{author}{\bibfnamefont{A.-L.} \bibnamefont{Barab\'{a}si}},
  \emph{\bibinfo{title}{Statistical mechanics of complex networks}}
  (\bibinfo{year}{2001}), \bibinfo{note}{e-print cond-mat/0106096}.

\bibitem[{\citenamefont{Dorogovtsev and
  Mendes}(2001{\natexlab{a}})}]{dorogorev}
\bibinfo{author}{\bibfnamefont{S.~N.} \bibnamefont{Dorogovtsev}}
  \bibnamefont{and} \bibinfo{author}{\bibfnamefont{J.~F.~F.}
  \bibnamefont{Mendes}}, \emph{\bibinfo{title}{Evolution of networks}}
  (\bibinfo{year}{2001}{\natexlab{a}}), \bibinfo{note}{e-print
  cond-mat/0106144}.

\bibitem[{nla()}]{nlanr}
\bibinfo{note}{The National Laboratory for Applied Network Research (NLANR),
  sponsored by the National Science Foundation, provides Internet routing
  related information based on border gateway protocol data (see
  http://moat.nlanr.net/)}.

\bibitem[{cai()}]{caida}
\bibinfo{note}{The Cooperative Association for Internet Data Analysis (CAIDA),
  located at the San Diego Supercomputer Center, provides measurements of
  Internet traffic metrics (see http://www.caida.org/home/)}.

\bibitem[{luc()}]{lucent}
\bibinfo{note}{B. Cheswick and H. Burch, Internet mapping project at Lucent
  Bell Labs. (http://www.cs.bell-labs.\-com/who\-/ches\-/map/)}.

\bibitem[{\citenamefont{Govindan and Reddy}(1997)}]{govindan97}
\bibinfo{author}{\bibfnamefont{R.}~\bibnamefont{Govindan}} \bibnamefont{and}
  \bibinfo{author}{\bibfnamefont{A.}~\bibnamefont{Reddy}}, in
  \emph{\bibinfo{booktitle}{Proceedings of IEEE INFOCOM'97}}
  (\bibinfo{year}{1997}), p. \bibinfo{pages}{850}.

\bibitem[{\citenamefont{Pansiot and Grad}(1998)}]{pansiot98}
\bibinfo{author}{\bibfnamefont{J.-J.} \bibnamefont{Pansiot}} \bibnamefont{and}
  \bibinfo{author}{\bibfnamefont{D.}~\bibnamefont{Grad}}, \bibinfo{journal}{ACM
  Comp. Comm. Rev.} \textbf{\bibinfo{volume}{28}}, \bibinfo{pages}{41}
  (\bibinfo{year}{1998}).

\bibitem[{\citenamefont{Faloutsos et~al.}(1999)\citenamefont{Faloutsos,
  Faloutsos, and Faloutsos}}]{falou99}
\bibinfo{author}{\bibfnamefont{M.}~\bibnamefont{Faloutsos}},
  \bibinfo{author}{\bibfnamefont{P.}~\bibnamefont{Faloutsos}},
  \bibnamefont{and}
  \bibinfo{author}{\bibfnamefont{C.}~\bibnamefont{Faloutsos}},
  \bibinfo{journal}{ACM SIGCOMM '99, Comput. Commun. Rev.}
  \textbf{\bibinfo{volume}{29}}, \bibinfo{pages}{251} (\bibinfo{year}{1999}).

\bibitem[{\citenamefont{Chou}(2000)}]{chou00}
\bibinfo{author}{\bibfnamefont{H.}~\bibnamefont{Chou}}, \emph{\bibinfo{title}{A
  note on power-laws of {I}nternet topology}} (\bibinfo{year}{2000}),
  \bibinfo{note}{e-print cs.NI/0012019}.

\bibitem[{\citenamefont{Caldarelli et~al.}(2000)\citenamefont{Caldarelli,
  Marchetti, and Pietronero}}]{calda00}
\bibinfo{author}{\bibfnamefont{G.}~\bibnamefont{Caldarelli}},
  \bibinfo{author}{\bibfnamefont{R.}~\bibnamefont{Marchetti}},
  \bibnamefont{and}
  \bibinfo{author}{\bibfnamefont{L.}~\bibnamefont{Pietronero}},
  \bibinfo{journal}{Europhys. Lett.} \textbf{\bibinfo{volume}{52}},
  \bibinfo{pages}{386} (\bibinfo{year}{2000}).

\bibitem[{\citenamefont{Yook et~al.}(2001)\citenamefont{Yook, Jeong, and
  Barab{\'a}si}}]{yook01}
\bibinfo{author}{\bibfnamefont{S.-H.} \bibnamefont{Yook}},
  \bibinfo{author}{\bibfnamefont{H.}~\bibnamefont{Jeong}}, \bibnamefont{and}
  \bibinfo{author}{\bibfnamefont{A.-L.} \bibnamefont{Barab{\'a}si}},
  \emph{\bibinfo{title}{Modeling the {I}nternet's large-scale topology}}
  (\bibinfo{year}{2001}), \bibinfo{note}{e-print cond-mat/0107417}.

\bibitem[{\citenamefont{Barab\'{a}si and Albert}(1999)}]{barab99}
\bibinfo{author}{\bibfnamefont{A.-L.} \bibnamefont{Barab\'{a}si}}
  \bibnamefont{and} \bibinfo{author}{\bibfnamefont{R.}~\bibnamefont{Albert}},
  \bibinfo{journal}{Science} \textbf{\bibinfo{volume}{286}},
  \bibinfo{pages}{509} (\bibinfo{year}{1999}).

\bibitem[{\citenamefont{Barab\'{a}si et~al.}(1999)\citenamefont{Barab\'{a}si,
  Albert, and Jeong}}]{barab992}
\bibinfo{author}{\bibfnamefont{A.-L.} \bibnamefont{Barab\'{a}si}},
  \bibinfo{author}{\bibfnamefont{R.}~\bibnamefont{Albert}}, \bibnamefont{and}
  \bibinfo{author}{\bibfnamefont{H.}~\bibnamefont{Jeong}},
  \bibinfo{journal}{Physica A} \textbf{\bibinfo{volume}{272}},
  \bibinfo{pages}{173} (\bibinfo{year}{1999}).

\bibitem[{\citenamefont{Doar and Leslie}(1993)}]{doar93}
\bibinfo{author}{\bibfnamefont{M.}~\bibnamefont{Doar}} \bibnamefont{and}
  \bibinfo{author}{\bibfnamefont{I.}~\bibnamefont{Leslie}}, in
  \emph{\bibinfo{booktitle}{Proceedings of IEEE INFOCOM'93}}
  (\bibinfo{year}{1993}), p.~\bibinfo{pages}{83}.

\bibitem[{\citenamefont{Floyd and Paxson}(2001)}]{paxson97}
\bibinfo{author}{\bibfnamefont{S.}~\bibnamefont{Floyd}} \bibnamefont{and}
  \bibinfo{author}{\bibfnamefont{V.}~\bibnamefont{Paxson}},
  \bibinfo{journal}{IEEE/ACM T. Network.} \textbf{\bibinfo{volume}{9}},
  \bibinfo{pages}{392} (\bibinfo{year}{2001}).

\bibitem[{\citenamefont{Adamic et~al.}(2001)\citenamefont{Adamic, Lukose,
  Puniyani, and Huberman}}]{adamic01}
\bibinfo{author}{\bibfnamefont{L.~A.} \bibnamefont{Adamic}},
  \bibinfo{author}{\bibfnamefont{R.~M.} \bibnamefont{Lukose}},
  \bibinfo{author}{\bibfnamefont{A.~R.} \bibnamefont{Puniyani}},
  \bibnamefont{and} \bibinfo{author}{\bibfnamefont{B.~A.}
  \bibnamefont{Huberman}}, \bibinfo{journal}{Phys. Rev. E}
  \textbf{\bibinfo{volume}{64}}, \bibinfo{pages}{046135}
  (\bibinfo{year}{2001}).

\bibitem[{\citenamefont{Puniyani et~al.}(2001)\citenamefont{Puniyani, Lukose,
  and Huberman}}]{puniyani01}
\bibinfo{author}{\bibfnamefont{A.~R.} \bibnamefont{Puniyani}},
  \bibinfo{author}{\bibfnamefont{R.~M.} \bibnamefont{Lukose}},
  \bibnamefont{and} \bibinfo{author}{\bibfnamefont{B.~A.}
  \bibnamefont{Huberman}}, \emph{\bibinfo{title}{Intentional walks on scale
  free small worlds}} (\bibinfo{year}{2001}), \bibinfo{note}{e-print
  cond-mat/0107212}.

\bibitem[{\citenamefont{Pastor-Satorras and Vespignani}(2001)}]{pv01a}
\bibinfo{author}{\bibfnamefont{R.}~\bibnamefont{Pastor-Satorras}}
  \bibnamefont{and}
  \bibinfo{author}{\bibfnamefont{A.}~\bibnamefont{Vespignani}},
  \bibinfo{journal}{Phys. Rev. Lett.} \textbf{\bibinfo{volume}{86}},
  \bibinfo{pages}{3200} (\bibinfo{year}{2001}).

\bibitem[{\citenamefont{Albert et~al.}(2000)\citenamefont{Albert, Jeong, and
  Barab\'{a}si}}]{barabasi00}
\bibinfo{author}{\bibfnamefont{R.~A.} \bibnamefont{Albert}},
  \bibinfo{author}{\bibfnamefont{H.}~\bibnamefont{Jeong}}, \bibnamefont{and}
  \bibinfo{author}{\bibfnamefont{A.-L.} \bibnamefont{Barab\'{a}si}},
  \bibinfo{journal}{Nature} \textbf{\bibinfo{volume}{406}},
  \bibinfo{pages}{378} (\bibinfo{year}{2000}).

\bibitem[{\citenamefont{Callaway et~al.}(2000)\citenamefont{Callaway, Newman,
  Strogatz, and Watts}}]{newman00}
\bibinfo{author}{\bibfnamefont{D.~S.} \bibnamefont{Callaway}},
  \bibinfo{author}{\bibfnamefont{M.~E.~J.} \bibnamefont{Newman}},
  \bibinfo{author}{\bibfnamefont{S.~H.} \bibnamefont{Strogatz}},
  \bibnamefont{and} \bibinfo{author}{\bibfnamefont{D.~J.} \bibnamefont{Watts}},
  \bibinfo{journal}{Phys. Rev. Lett.} \textbf{\bibinfo{volume}{85}},
  \bibinfo{pages}{5468} (\bibinfo{year}{2000}).

\bibitem[{\citenamefont{Cohen et~al.}(2001)\citenamefont{Cohen, Erez,
  {ben-Avraham}, and Havlin}}]{havlin01}
\bibinfo{author}{\bibfnamefont{R.}~\bibnamefont{Cohen}},
  \bibinfo{author}{\bibfnamefont{K.}~\bibnamefont{Erez}},
  \bibinfo{author}{\bibfnamefont{D.}~\bibnamefont{{ben-Avraham}}},
  \bibnamefont{and} \bibinfo{author}{\bibfnamefont{S.}~\bibnamefont{Havlin}},
  \bibinfo{journal}{Phys. Rev. Lett.} \textbf{\bibinfo{volume}{86}},
  \bibinfo{pages}{3682} (\bibinfo{year}{2001}).

\bibitem[{\citenamefont{Watts and Strogatz}(1998)}]{watts98}
\bibinfo{author}{\bibfnamefont{D.~J.} \bibnamefont{Watts}} \bibnamefont{and}
  \bibinfo{author}{\bibfnamefont{S.~H.} \bibnamefont{Strogatz}},
  \bibinfo{journal}{Nature} \textbf{\bibinfo{volume}{393}},
  \bibinfo{pages}{440} (\bibinfo{year}{1998}).

\bibitem[{\citenamefont{Pastor-Satorras
  et~al.}(2001)\citenamefont{Pastor-Satorras, V{\'a}zquez, and
  Vespignani}}]{alexei}
\bibinfo{author}{\bibfnamefont{R.}~\bibnamefont{Pastor-Satorras}},
  \bibinfo{author}{\bibfnamefont{A.}~\bibnamefont{V{\'a}zquez}},
  \bibnamefont{and}
  \bibinfo{author}{\bibfnamefont{A.}~\bibnamefont{Vespignani}},
  \bibinfo{journal}{Phys. Rev. Lett.} \textbf{\bibinfo{volume}{87}},
  \bibinfo{pages}{258701} (\bibinfo{year}{2001}).

\bibitem[{mer()}]{mercator}
\bibinfo{note}{Mapping the Internet within the SCAN project at the information
  Sciences Institute (http://www.\-isi.\-edu\-/div7\-/scan/)}.

\bibitem[{\citenamefont{Bollob{\'a}s}(1985)}]{bollobas}
\bibinfo{author}{\bibfnamefont{B.}~\bibnamefont{Bollob{\'a}s}},
  \emph{\bibinfo{title}{Random Graphs}} (\bibinfo{publisher}{Academic Press},
  \bibinfo{address}{London}, \bibinfo{year}{1985}).

\bibitem[{\citenamefont{Watts}(1999)}]{watts99}
\bibinfo{author}{\bibfnamefont{D.~J.} \bibnamefont{Watts}},
  \emph{\bibinfo{title}{Small worlds: The dynamics of networks between order
  and randomness}} (\bibinfo{publisher}{Princeton University Press},
  \bibinfo{address}{New Jersey}, \bibinfo{year}{1999}).

\bibitem[{\citenamefont{Newman}(2001)}]{newman01b}
\bibinfo{author}{\bibfnamefont{M.~E.~J.} \bibnamefont{Newman}},
  \bibinfo{journal}{Phys. Rev. E} \textbf{\bibinfo{volume}{64}},
  \bibinfo{pages}{016132} (\bibinfo{year}{2001}).

\bibitem[{\citenamefont{Goh et~al.}(2001)\citenamefont{Goh, Kahng, and
  Kim}}]{goh01}
\bibinfo{author}{\bibfnamefont{K.-I.} \bibnamefont{Goh}},
  \bibinfo{author}{\bibfnamefont{B.}~\bibnamefont{Kahng}}, \bibnamefont{and}
  \bibinfo{author}{\bibfnamefont{D.}~\bibnamefont{Kim}},
  \bibinfo{journal}{Phys. Rev. Lett.}  (\bibinfo{year}{2001}).

\bibitem[{\citenamefont{Krapivsky et~al.}(2000)\citenamefont{Krapivsky, Redner,
  and Leyvraz}}]{krap00}
\bibinfo{author}{\bibfnamefont{P.~L.} \bibnamefont{Krapivsky}},
  \bibinfo{author}{\bibfnamefont{S.}~\bibnamefont{Redner}}, \bibnamefont{and}
  \bibinfo{author}{\bibfnamefont{F.}~\bibnamefont{Leyvraz}},
  \bibinfo{journal}{Phys. Rev. Lett.} \textbf{\bibinfo{volume}{85}},
  \bibinfo{pages}{4629} (\bibinfo{year}{2000}).

\bibitem[{\citenamefont{Krapivsky and Redner}(2001)}]{krap00b}
\bibinfo{author}{\bibfnamefont{P.~L.} \bibnamefont{Krapivsky}}
  \bibnamefont{and} \bibinfo{author}{\bibfnamefont{S.}~\bibnamefont{Redner}},
  \bibinfo{journal}{Phys. Rev. E} \textbf{\bibinfo{volume}{63}},
  \bibinfo{pages}{066123} (\bibinfo{year}{2001}).

\bibitem[{\citenamefont{Jeong et~al.}(2001)\citenamefont{Jeong, N\'{e}da, and
  Barab\'{a}si}}]{barabba01b}
\bibinfo{author}{\bibfnamefont{H.}~\bibnamefont{Jeong}},
  \bibinfo{author}{\bibfnamefont{Z.}~\bibnamefont{N\'{e}da}}, \bibnamefont{and}
  \bibinfo{author}{\bibfnamefont{A.-L.} \bibnamefont{Barab\'{a}si}},
  \emph{\bibinfo{title}{Measuring preferential attachment for evolving
  networks}} (\bibinfo{year}{2001}), \bibinfo{note}{e-print cond-mat/0104131}.

\bibitem[{\citenamefont{Erd\"{o}s and R\'{e}nyi}(1960)}]{erdos60}
\bibinfo{author}{\bibfnamefont{P.}~\bibnamefont{Erd\"{o}s}} \bibnamefont{and}
  \bibinfo{author}{\bibfnamefont{P.}~\bibnamefont{R\'{e}nyi}},
  \bibinfo{journal}{Publ. Math. Inst. Hung. Acad. Sci.}
  \textbf{\bibinfo{volume}{5}}, \bibinfo{pages}{17} (\bibinfo{year}{1960}).

\bibitem[{\citenamefont{Zegura et~al.}(1997)\citenamefont{Zegura, Calvert, and
  Donahoo}}]{zeg}
\bibinfo{author}{\bibfnamefont{E.~W.} \bibnamefont{Zegura}},
  \bibinfo{author}{\bibfnamefont{K.}~\bibnamefont{Calvert}}, \bibnamefont{and}
  \bibinfo{author}{\bibfnamefont{M.~J.} \bibnamefont{Donahoo}},
  \bibinfo{journal}{IEEE/ACM T. Network.} \textbf{\bibinfo{volume}{5}},
  \bibinfo{pages}{770} (\bibinfo{year}{1997}), \bibinfo{note}{and references
  therein}.

\bibitem[{\citenamefont{Dorogovtsev
  et~al.}(2000{\natexlab{a}})\citenamefont{Dorogovtsev, J.~F. F.~Mendes, and
  Samukhin}}]{mendes99}
\bibinfo{author}{\bibfnamefont{S.~N.} \bibnamefont{Dorogovtsev}},
  \bibinfo{author}{\bibfnamefont{J.}~\bibnamefont{J.~F. F.~Mendes}},
  \bibnamefont{and} \bibinfo{author}{\bibfnamefont{A.~N.}
  \bibnamefont{Samukhin}}, \bibinfo{journal}{Phys. Rev. Lett.}
  \textbf{\bibinfo{volume}{85}}, \bibinfo{pages}{4633}
  (\bibinfo{year}{2000}{\natexlab{a}}).

\bibitem[{\citenamefont{Simon}(1955)}]{simon55}
\bibinfo{author}{\bibfnamefont{H.~A.} \bibnamefont{Simon}},
  \bibinfo{journal}{Biometrika} \textbf{\bibinfo{volume}{42}},
  \bibinfo{pages}{425} (\bibinfo{year}{1955}).

\bibitem[{\citenamefont{Bornholdt and Ebel}(2001)}]{bornholdt01}
\bibinfo{author}{\bibfnamefont{S.}~\bibnamefont{Bornholdt}} \bibnamefont{and}
  \bibinfo{author}{\bibfnamefont{H.}~\bibnamefont{Ebel}},
  \bibinfo{journal}{Phys. Rev. E} \textbf{\bibinfo{volume}{64}},
  \bibinfo{pages}{035104} (\bibinfo{year}{2001}).

\bibitem[{\citenamefont{Dorogovtsev
  et~al.}(2000{\natexlab{b}})\citenamefont{Dorogovtsev, Mendes, and
  Samukhin}}]{dmscomment}
\bibinfo{author}{\bibfnamefont{S.~N.} \bibnamefont{Dorogovtsev}},
  \bibinfo{author}{\bibfnamefont{J.~F.~F.} \bibnamefont{Mendes}},
  \bibnamefont{and} \bibinfo{author}{\bibfnamefont{A.~N.}
  \bibnamefont{Samukhin}}, \emph{\bibinfo{title}{{WWW} and {I}nternet models
  from 1955 till our days and the ``popularity is attractive'' principle}}
  (\bibinfo{year}{2000}{\natexlab{b}}), \bibinfo{note}{e-print
  cond-mat/0009090}.

\bibitem[{\citenamefont{Albert and Barab{\'a}si}(2000)}]{albert00}
\bibinfo{author}{\bibfnamefont{R.}~\bibnamefont{Albert}} \bibnamefont{and}
  \bibinfo{author}{\bibfnamefont{A.-L.} \bibnamefont{Barab{\'a}si}},
  \bibinfo{journal}{Phys. Rev. Lett.} \textbf{\bibinfo{volume}{85}},
  \bibinfo{pages}{5234} (\bibinfo{year}{2000}).

\bibitem[{\citenamefont{Medina et~al.}(2000)\citenamefont{Medina, Matt, and
  Byers}}]{medina}
\bibinfo{author}{\bibfnamefont{A.}~\bibnamefont{Medina}},
  \bibinfo{author}{\bibfnamefont{I.}~\bibnamefont{Matt}}, \bibnamefont{and}
  \bibinfo{author}{\bibfnamefont{J.}~\bibnamefont{Byers}},
  \bibinfo{journal}{Comput. Commun. Rev.} \textbf{\bibinfo{volume}{30}},
  \bibinfo{pages}{18} (\bibinfo{year}{2000}).

\bibitem[{\citenamefont{Dorogovtsev and Mendes}(2000)}]{mendes00}
\bibinfo{author}{\bibfnamefont{S.~N.} \bibnamefont{Dorogovtsev}}
  \bibnamefont{and} \bibinfo{author}{\bibfnamefont{J.~F.~F.}
  \bibnamefont{Mendes}}, \bibinfo{journal}{Europhys. Lett.}
  \textbf{\bibinfo{volume}{52}}, \bibinfo{pages}{33} (\bibinfo{year}{2000}).

\bibitem[{\citenamefont{Dorogovtsev and Mendes}(2001{\natexlab{b}})}]{mendes01}
\bibinfo{author}{\bibfnamefont{S.~N.} \bibnamefont{Dorogovtsev}}
  \bibnamefont{and} \bibinfo{author}{\bibfnamefont{J.~F.~F.}
  \bibnamefont{Mendes}}, \bibinfo{journal}{Phys. Rev. E}
  \textbf{\bibinfo{volume}{63}}, \bibinfo{pages}{025101}
  (\bibinfo{year}{2001}{\natexlab{b}}).

\bibitem[{\citenamefont{Molloy and Reed}(1995)}]{molloy95}
\bibinfo{author}{\bibfnamefont{M.}~\bibnamefont{Molloy}} \bibnamefont{and}
  \bibinfo{author}{\bibfnamefont{B.}~\bibnamefont{Reed}},
  \bibinfo{journal}{Random Structures and Algorithms}
  \textbf{\bibinfo{volume}{6}}, \bibinfo{pages}{161} (\bibinfo{year}{1995}).

\bibitem[{\citenamefont{Molloy and Reed}(1998)}]{molloy98}
\bibinfo{author}{\bibfnamefont{M.}~\bibnamefont{Molloy}} \bibnamefont{and}
  \bibinfo{author}{\bibfnamefont{B.}~\bibnamefont{Reed}},
  \bibinfo{journal}{Combinatorics, Probability, and Computing}
  \textbf{\bibinfo{volume}{7}}, \bibinfo{pages}{295} (\bibinfo{year}{1998}).

\bibitem[{\citenamefont{Bianconi and Barab{\'a}si}(2001)}]{bianconi01}
\bibinfo{author}{\bibfnamefont{G.}~\bibnamefont{Bianconi}} \bibnamefont{and}
  \bibinfo{author}{\bibfnamefont{A.-L.} \bibnamefont{Barab{\'a}si}},
  \bibinfo{journal}{Europhys. Lett} \textbf{\bibinfo{volume}{54}},
  \bibinfo{pages}{436} (\bibinfo{year}{2001}).

\bibitem[{\citenamefont{Newman et~al.}(2001)\citenamefont{Newman, Strogatz, and
  Watts}}]{newman01c}
\bibinfo{author}{\bibfnamefont{M.~E.~J.} \bibnamefont{Newman}},
  \bibinfo{author}{\bibfnamefont{S.~H.} \bibnamefont{Strogatz}},
  \bibnamefont{and} \bibinfo{author}{\bibfnamefont{D.~J.} \bibnamefont{Watts}},
  \bibinfo{journal}{Phys. Rev. E} \textbf{\bibinfo{volume}{64}},
  \bibinfo{pages}{026118} (\bibinfo{year}{2001}).

\end{thebibliography}
\end{document}